\documentclass[lettersize,journal]{IEEEtran}
\IEEEoverridecommandlockouts
\usepackage{cite}
\usepackage{amsmath,amssymb,amsfonts}
\usepackage{algorithmic}
\usepackage{graphicx}
\usepackage{textcomp}
\usepackage{xcolor}
\usepackage{microtype}
\usepackage{algorithm}
\usepackage{array}
\usepackage{microtype}
\usepackage{stfloats}
\usepackage{url}
\usepackage{verbatim}
\usepackage{graphicx}
\usepackage{cite}
\usepackage{subfigure}
\usepackage{hyperref} 

\hypersetup{
    colorlinks=true,   
    linkcolor=blue,    
    citecolor=blue,    
    urlcolor=blue      
}
\usepackage{booktabs}

\def\BibTeX{{\rm B\kern-.05em{\sc i\kern-.025em b}\kern-.08em
    T\kern-.1667em\lower.7ex\hbox{E}\kern-.125emX}}
\begin{document}

\title{FDMA-Based Passive Multiple Users SWIPT Utilizing Resonant Beams\\
}

\author{Yixuan Guo, Mingliang Xiong,~\IEEEmembership{Member,~IEEE,} Wen Fang,~\IEEEmembership{Member,~IEEE,} Qingwei Jiang, \\Qingwen Liu,~\IEEEmembership{Senior Member,~IEEE,} and 
Gang Yan,~\IEEEmembership{Member,~IEEE}
\thanks{Y. Guo is with the Shanghai Research Institute for Intelligent Autonomous Systems, Tongji University, Shanghai 201210, China 
(e-mail: guoyixuan@tongji.edu.cn).

M. Xiong, Q. Jiang, and Q. Liu are with the School of Computer Science and Technology, Tongji University, Shanghai 201804, China (e-mail:  mlx@tongji.edu.com; jiangqw@tongji.edu.cn; qliu@tongji.edu.cn).

W. Fang is with the College of Electronic and Information Engineering, Tongji University, Shanghai 201804, China (e-mail: wen.fang@tongji.edu.cn).

G. Yan is with the School of Physics Science and Engineering, Tongji University, Shanghai 200092, China (e-mail: gyan@tongji.edu.cn).
	}
}

\maketitle

\begin{abstract}
The rapid development of IoT technology has led to a shortage of spectrum resources and energy, giving rise to simultaneous wireless information and power transfer (SWIPT) technology. However, traditional multiple input multiple output (MIMO)-based SWIPT faces challenges in target detection. We have designed a passive multi-user resonant beam system (MU-RBS) that can achieve efficient power transfer and communication through adaptive beam alignment. 
The frequency division multiple access (FDMA) is employed in the downlink (DL) channel, while frequency conversion is utilized in the uplink (UL) channel to avoid echo interference and co-channel interference, and the system architecture design and corresponding mathematical model are presented.
The simulation results show that MU-RBS can achieve adaptive beam-forming without the target transmitting pilot signals, has high directivity, and as the number of iterations increases, the power transmission efficiency, signal-to-noise ratio and spectral efficiency of the UL and DL are continuously optimized until the system reaches the optimal state. 
\end{abstract}


\section{Introduction}
With the rapid development of Internet of Things (IoT) technology and artificial intelligence, the number of IoT devices worldwide has surged, leading to a tight competition for spectrum resources and energy \cite{wikstrom2020challenges}\cite{shinohara2020trends}. This has prompted the next generation of wireless communication technologies to not only focus on improving data rates but also emphasize energy efficiency to provide strong support for applications in smart cities, smart transportation, and other fields.

Simultaneous wireless information and power transfer (SWIPT) technology is an emerging network technology that integrates wireless information transfer (WIT) and wireless power transfer (WPT) into the same wireless channel, achieving efficient utilization of wireless resources \cite{9675011},\cite{an2021energy}. In a SWIPT system, network nodes can not only transmit data but also alleviate battery life concerns by providing charging services to devices.

In traditional SWIPT systems, power is typically transmitted to the receiver via a fixed transmitting antenna, which results in low transmission efficiency, especially when facing multi-directional transmission \cite{perera2017simultaneous}. Multiple input multiple output (MIMO) technology, widely used in wireless communication, has recently been introduced to WPT. It can be extended to inductive coupling WPT \cite{yang2017magnetic} and radio frequency (RF) based WPT \cite{yang2018wireless}, where multiple antenna arrays are used to achieve spatial reuse of information and energy, thus enhancing communication coverage and power transmission efficiency. Coupled with beamforming technology, energy wastage can be further reduced \cite{moghadam2018energy}.

However, in MIMO-based SWIPT, target detection is a critical challenge because the direction of arrival (DOA) needs to be estimated based on the received pilot signals before beamforming, and then the phase shifters are adjusted to form the optimal beam. This process is complex and requires the target to have the energy to actively transmit signals. In schemes \cite{zhang2018wireless} and \cite{mitani2019experimental}, after the target receives the electromagnetic waves radiated by the base station (BS), harmonics generated by rectifiers replace the pilot signals, which can be used in passive scenarios. However, it is not straightforward to achieve in multi-path environments.

The authors of \cite{zeng2017communications} proposed using the retro-directive beamforming technique to achieve multi-target WPT. The BS conjugates and amplifies the total signal from the uplink (UL) through a retro-directive array (RDA) to improve the transmission efficiency of the downlink (DL) and reduce system complexity. However, due to the round-trip signal propagation loss, there is a ``dual near far'' problem in multi-target scenarios \cite{ju2013throughput}. Therefore, \cite{lee2017retrodirective} proposed a distributed UL power update algorithm to adjust the UL power of users. Although the design requires users to actively send signals, the potential of the RDA has been demonstrated.
Passive beamforming is one of the advantages of intelligent reflecting surfaces (IRS) \cite{wu2019towards},\cite{wu2021intelligent},\cite{gong2020toward}. The authors of \cite{gao2022beamforming} and \cite{wu2020joint} propose IRS-based SWIPT architectures, where both the BS and user equipment (UE) can perform active beamforming to obtain channel state information (CSI). The IRS can align the beam to the UE in a passive state by acquiring CSI from the BS. However, if the UE is also passive,  CSI acquisition becomes difficult, which will increase the complexity of system implementation. \cite{10924145} proposes spatially separated laser cavities equipped at both the BS and UE, eliminating the need to track the UE, as the IRS can adaptively achieve beam alignment. However, optical lenses are generally unsuitable for low-power IoT devices and suffer from low photoelectric conversion efficiency. 
In \cite{jiang2025single}, both the BS and UE are equipped with a RDA based on heterodyne mixing. This setup enables self-aligned WPT through round-trip electromagnetic wave resonance, eliminating the need for DOA estimation. However, the cyclic oscillation of electromagnetic waves with the same round-trip frequency causes echo interference, making it unsuitable for WIT systems. The authors of \cite{xia2024millimeter} also employ a heterodyne-mixing RDA, addressing the echo interference by loading different frequencies at the transmitter to achieve frequency separation for UL and DL, thus realizing SWIPT. However, this frequency separation method lacks flexibility, and the heterodyne mixing results in low isolation between RF and intermediate frequencies. Additionally, the aforementioned RDA systems require the UE to transmit a pilot signal to activate the RDA at the BS for direction retracing, and they are only applicable to single-user scenarios.

\begin{figure}[!t]
\centering
\includegraphics[width=0.8\linewidth]{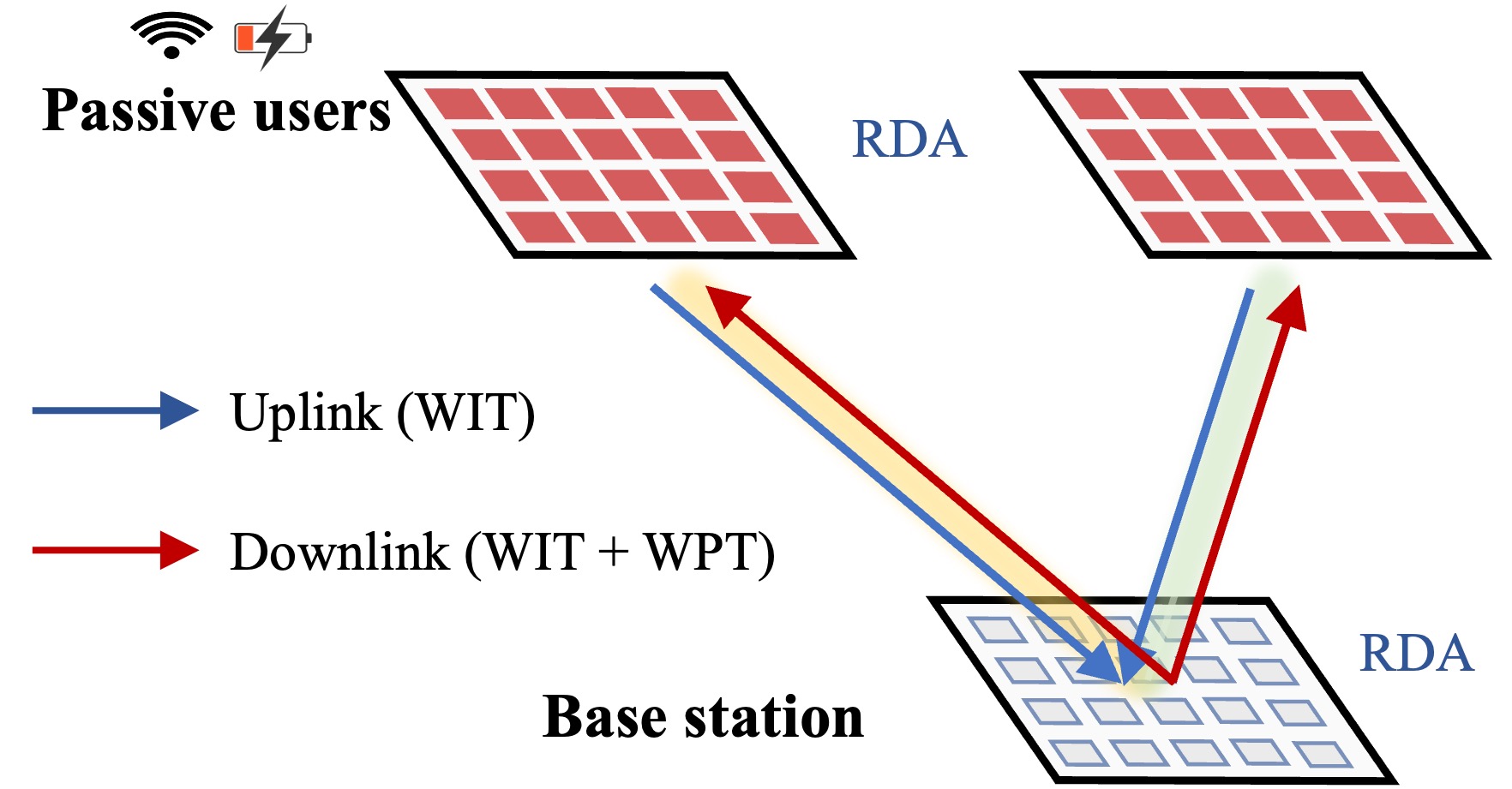}
\caption{Application of SWIPT based on resonant beam in passive IoT devices.}
\label{fig1}
\end{figure}

To achieve multi-user SWIPT in passive scenarios, the system design must minimize the complexity of UE detection without requiring the UE to transmit pilot signals, while effectively addressing co-channel interference among users. Therefore, this paper proposes a multi-user resonant beam system (MU-RBS), as illustrated in Fig.~\ref{fig1}. In this system, both the BS and each UE are equipped with the RDAs. In the initial stage of the system, the BS broadcasts electromagnetic waves carrying both power and information. Upon receiving the signals, the UEs store some power to drive RDA operation and return the remaining power to the BS. After receiving the return signal, the RDA of BS dynamically adjusts the beam direction and traces back to the corresponding UE based on the position and feedback information of each UE, thereby achieving adaptive beam alignment. Through multiple iterations, the round-trip waves between BS and UE can establish resonance, ultimately enabling high-efficiency power transfer and high spectral efficiency communication. Unlike \cite{jiang2025single} and \cite{xia2024millimeter}, the proposed system in this paper incorporates phase-locked loop (PLL) which can generate new frequencies based on the input frequency, effectively suppressing echo interference. Additionally, the system employs frequency division multiple access (FDMA) to mitigate co-channel interference among different links, realizing SWIPT in multi-user scenarios.
The main contributions of the paper are as follows:
\begin{itemize}
\item To address the demand for SWIPT in passive multi-user scenarios, a novel RF system architecture based on RDA is proposed. This system achieves adaptive beamforming through resonance characteristics without relying on pilot signal transmission from the UE or complex UE detection mechanisms, significantly reducing system control complexity while supporting multi-user access.

\item By incorporating PLL, the system accomplishes frequency switching while achieving phase conjugation of input/output electromagnetic waves, effectively mitigating echo interference caused by the cyclic oscillation of round-trip electromagnetic waves. Furthermore, FDMA is employed to avoid co-channel interference in multi-user scenarios.

\item We established mathematical models fo WPT and WIT of the proposed system, and derived closed form expressions for power transmission efficiency and spectral efficiency. Numerical simulations have verified that the system can achieve efficient power transfer and high spectral efficiency communication while enabling adaptive beamforming.

\end{itemize}

The remainder of this paper is organized as follows: Section \ref{sec2} describes the structure of the BS and UE transceivers, as well as the principle of resonance formation; Section \ref{sec3} presents the mathematical models for wireless power transfer and wireless information transfer; Section \ref{sec4} conducts a simulation analysis of the power transfer and communication performance of the proposed MU-RBS; finally, Section \ref{sec5} concludes this research work.

\label{sec2}
\section{System Desgin}

\begin{figure*}[ht]
\centering
\includegraphics[width=0.9\linewidth]{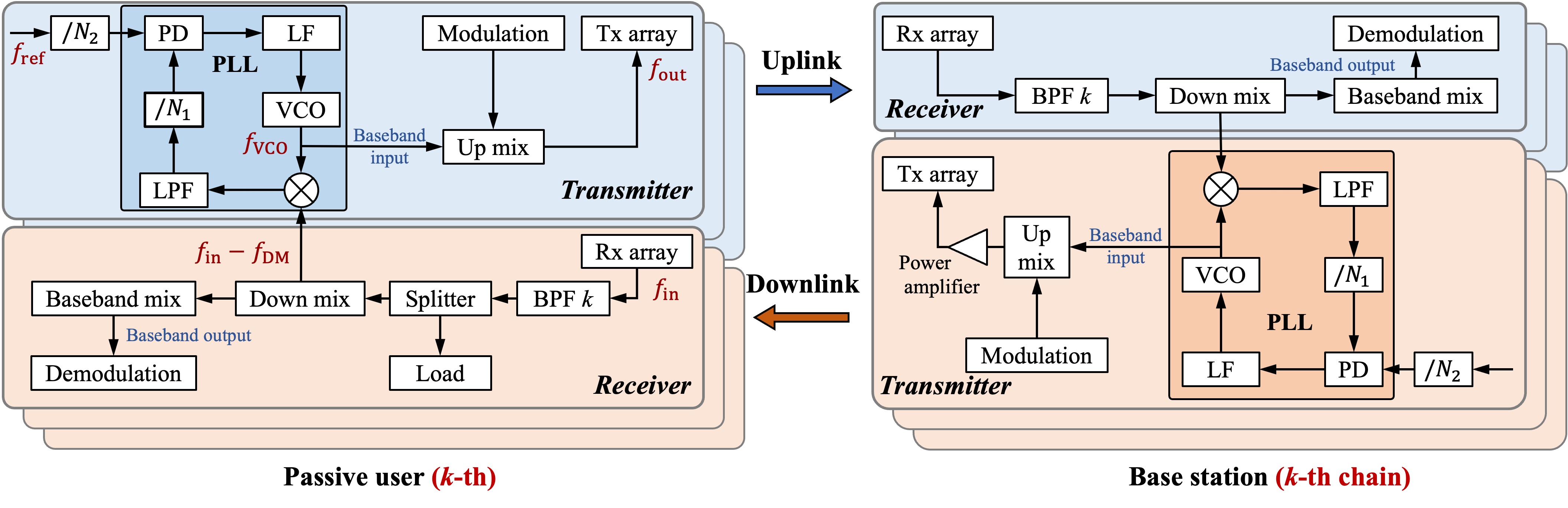}
  \caption{Structure of the base station and the $k$-th passive user equipment transceiver.}
\label{system}
\end{figure*}

In this section, we present the system design of MU-RBS. First, we introduce the overall FDMA-based RBS architecture and elaborate on the principles of phase conjugation and frequency conversion. 
 Subsequently, we specify the requirements for parallel uplink-downlink transmission and analyze the system stabilization criteria.

\subsection{Structural design of the system}

In the MU-RBS, FDMA technology is employed to ensure that the communication links between each passive UE and the BS do not interfere with each other. Assuming there are $K$ UEs in the system, the BS needs to be equipped with at least $K$ RF chains to meet the communication requirements of each UE. Additionally, to achieve resonance and avoid interference between the UL and DL, the arrays of both the UEs and the BS must possess the capability of phase conjugation and frequency conversion. To provide a clearer description of the system's working principle, Fig.~\ref{system} illustrates the transceiver structure of the $k$-th UE and the BS.

Taking the $k$-th UE as an example, the BS first transmits an electromagnetic wave to the UE. The UE then filters the received signal through a bandpass filter (BPF), retaining only the supported frequency band with frequency $f_\text{in}$ and phase $\varphi_\text{in}$. 
Subsequently, the signal passes through a power splitter, where a portion of the signal is diverted for power harvesting while the remaining portion is used for signal processing.

In the processing branch, the signal undergoes first-stage down mix to $f_\text{in} - f_\text{dm}$, followed by a second split where one portion undergoes second-stage baseband mix to near-baseband frequency $(f_\text{in} - f_\text{dm} - f_\text{bm} \approx 0)$ for demodulation, while the other branch enters a phase-locked loop (PLL) for phase conjugation and frequency conversion to generate the backscatter signal.

The signal entering the PLL first passes through a frequency divider 1 to reduce the frequency to $(f_\text{in}-f_\text{dm})/N_1$. Then it enters a phase detector (PD), where it is compared with a reference signal $f_\text{ref}/N_2$ that has also been divided. Based on the phase difference, a voltage is generated to drive a voltage-controlled oscillator (VCO), producing a signal with frequency $f_\text{vco}$. This signal is then mixed with the input signal and passed through a low-pass filter (LPF) to remove high frequency components. The filtered signal is again compared to the reference signal, forming a closed-loop control system. A loop filter is used to eliminate unnecessary high frequency signals and fluctuations generated during the process. Finally, when the phase difference approaches zero, the system reaches a locked state and the output signal is generated \cite{buchanan2011high}, \cite{1456254}. At this point, the relationship between the PD's output and input signals can be characterized by their frequency and phase correspondence, expressed as

\begin{equation}
    \frac{f_\text{ref}}{N_2}=\frac{f_\text{in}+f_\text{vco}-f_\text{dm}}{N_1},
\end{equation}
\begin{equation}
    \frac{\varphi_\text{ref}}{N_2}=\frac{\varphi_\text{in}+\varphi_\text{vco}-\varphi_\text{dm}}{N_1}.
\end{equation}

Assuming all oscillators and the reference signal have zero initial phase, the phase conjugation condition is satisfied
\begin{equation}
    \varphi_\text{vco}+\varphi_\text{in}=0.
\end{equation}

The VCO output undergoes upconversion with the modulation signal to generate the backscattered signal, resulting in a final output frequency expressed as
\begin{align}
\nonumber f_\text{out}&=f_\text{vco}+f_\text{um}\\&=\frac{N_1}{N_2}f_\text{ref}+f_\text{dm}+f_\text{um}-f_\text{in},
\end{align}
where $\frac{N_1}{N_2}$ is the frequency divider ratio in the PLL. This demonstrates programmable frequency control via the reference signal and dividers.

The signal sent by UE is received and decoded by the $k$-th RF link of BS.
BS also needs to perform phase conjugation and frequency conversion on the received signal to ensure that the output signal frequency matches the receiving frequency of the corresponding UE. The main difference in structure between BS and UE is that BS must support the access of multiple UEs simultaneously and must be equipped with a power amplifier (PA) to compensate for the energy loss generated by the signal during each cycle, thus ensuring the stability of the communication link and the efficiency of energy transmission.

\subsection{Principle of resonance formation}

\begin{figure}[ht]
\centering
\includegraphics[width=0.9\linewidth]{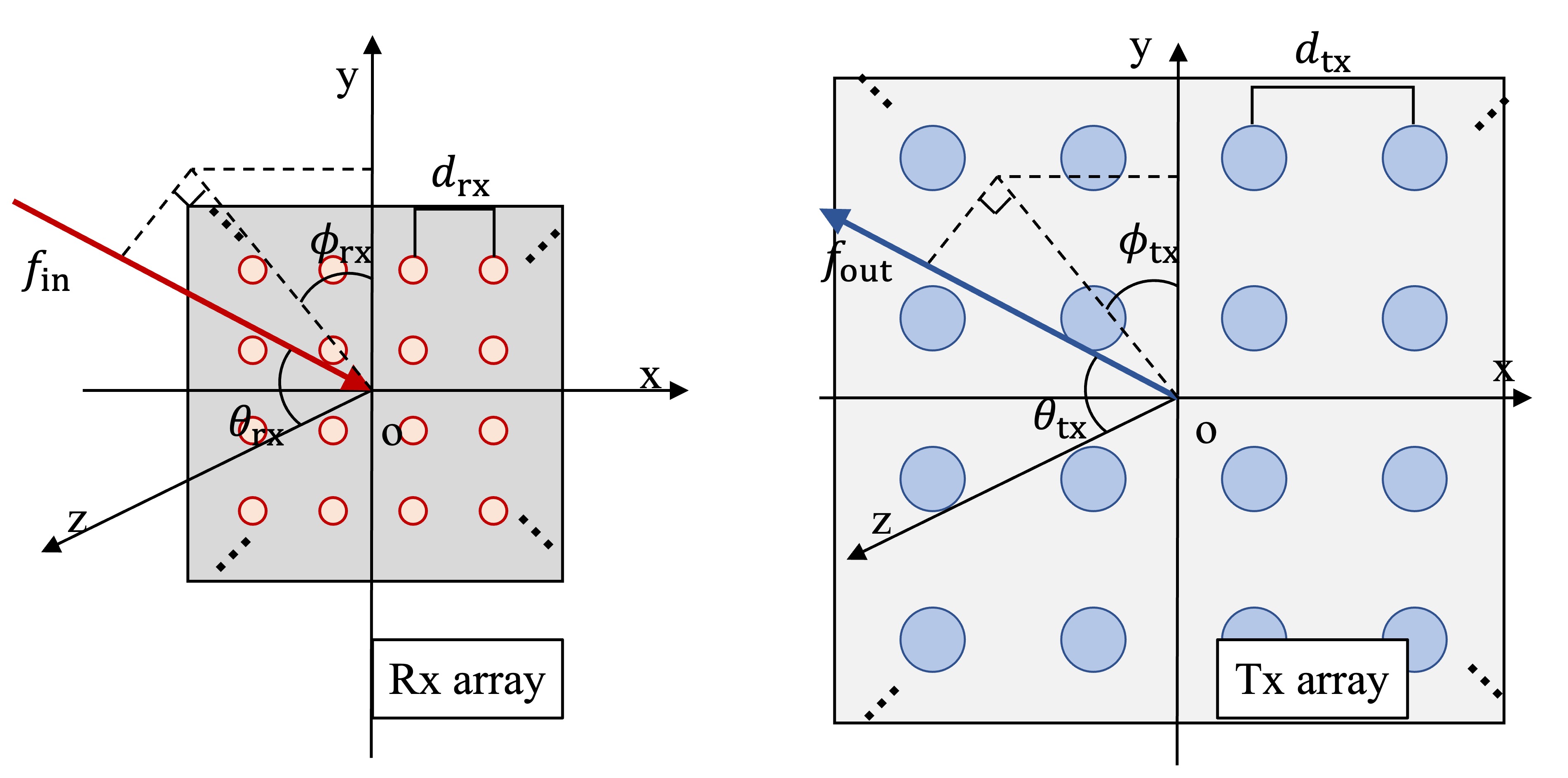}
  \caption{The electromagnetic wave original path feedback principle.}
\label{doa}
\end{figure}
In conventional RDAs, since the transmitted and received signals share the same frequency, both transmission and reception functions can be realized using a single antenna array.
To achieve frequency-separated Tx/Rx arrays, the inherent physical isolation between different frequency bands can be utilized to prevent direct coupling between signals in the same frequency range  \cite{9921326}.
Common array arrangement methods primarily include stacked configuration \cite{9955558} and interleaved planar configuration \cite{sandhu2016radiating},\cite{xu2019research}. 
In this paper, we adopt the interleaved planar arrangement, assuming co-located Tx/Rx array centers.

Due to the difference between the input frequency and output frequency of UEs and BS, in order to ensure that the output signal is parallel to the transmission direction of the received signal, it is necessary not only to control the initial phase of the transmitted signal to be opposite to the phase of the received signal but also to ensure that the phase difference between adjacent elements of the transmitting and receiving arrays is equal, that is,

\begin{align}
\nonumber&\frac{2\pi }{\lambda_\text{in}}  d_\text{rx}\sin\theta_\text{rx} \left(\sin\phi_\text{rx} + \cos\phi_\text{rx}\right) \\ = &\frac{2\pi}{\lambda_\text{out}} d_\text{tx}\sin\theta_\text{tx} \left(\sin\phi_\text{tx} + \cos\phi_\text{tx}\right)
\end{align}
where $\lambda_\text{in}$ and $\lambda_\text{out}$ are the wavelengths of the input and output signals, respectively. $d_\text{rx}$ and $d_\text{tx}$ are the element spacings of the receiving and transmitting arrays, respectively. $\theta_\text{rx}$ and $\phi_\text{rx}$ represent the elevation and azimuth angles of the incoming signal at the receiving array, while $\theta_\text{tx}$ and $\phi_\text{tx}$ represent the elevation and azimuth angles of the outgoing signal at the transmitting array. As shown as Fig.~\ref{doa}, when the received and transmitted signals are parallel, $\theta_\text{tx} = \theta_\text{rx}$ and $\phi_\text{tx} = \phi_\text{rx}$, the above equation simplifies to

\begin{equation}
\frac{d_\text{rx}}{\lambda_\text{in}} = \frac{d_\text{tx}}{\lambda_\text{out}}
\end{equation}
which implies that the output signal can retrace the original path when the element spacing of the receiving and transmitting arrays satisfies the following relationship \cite{guo2025integrated}.

\begin{equation}
\frac{d_\text{rx}}{d_\text{tx}} = \frac{\lambda_\text{in}}{\lambda_\text{out}}
\label{77}
\end{equation}

In MU-RBS, the amplifier at BS is the only power source. It needs to compensate for the energy loss during the continuous directional propagation of electromagnetic waves between the BS and each UE, as well as the energy stored by the UE. Therefore, the loss of the system can be evaluated based on the output power of the BS.

During the initial iteration process, due to phase misalignment, the propagation direction of the round-trip electromagnetic wave deviates, leading to energy dispersion in the sidelobe region and a significant increase in system loss. The loss coefficient $L(i)$ can be defined as

\begin{equation}
L(i) = \frac{\mathbf{P}_{b,k}^\text{r}(i)}{\mathbf{P}_{b,k}^\text{t}(i)}
\end{equation}
where $\mathbf{P}_{b,k}^\text{t}(i)$ represents the electromagnetic wave power radiated by the BS to the $k$-th UE during the $i$-th iteration, and $\mathbf{P}_{b,k}^\text{r}(i)$ denotes the electromagnetic wave power received by the BS from the $k$-th UE during the $i$-th iteration.

In the $i$-th iteration, the system gain $G(i)$ can be defined as

\begin{equation}
G(i) = \frac{\mathbf{P}_{b,k}^\text{t}(i)}{\mathbf{P}_{b,k}^\text{r}(i-1)}
\end{equation}

As the number of iterations increases, the system gradually converges and the gains and losses satisfy the following relationship

\begin{equation}
\lim\limits_{i\rightarrow\infty} G(i) L(i) = 1
\end{equation}

At this point, the gain fully compensates for the loss and the system reaches steady-state equilibrium.

From the physical mechanism perspective, during the iteration process, phase-misaligned electromagnetic waves undergo mutual attenuation due to interference effects, while phase-aligned electromagnetic waves form a standing wave through coherent superposition, thereby achieving a stable resonance state. In this state, the spatial distribution of the electromagnetic field reaches a steady configuration and self-replicates in subsequent iterations \cite{10636970},\cite{guo2024resonant}. At this point, the UL and DL signals of the system exhibit significant directionality, with energy concentrated in the main lobe region, which can be regarded as the realization of adaptive beamforming.
\section{Mathematical Analysis}
\label{sec3}

In this section, we establishe the theoretical SWIPT model for MU-RBS. First, we develop a bidirectional power transfer model between the BS and $K$ UEs, deriving analytical expressions for both DL and UL power transmission efficiency. Second, we construct the MU-RBS's channel model and derive closed-form expressions for the spectral efficiency of DL and UL transmissions.

\subsection{Wireless power transmission model}
We consider a SWIPT system with MIMO, where a BS equipped with $M$ antennas serves $K$ UEs equipped with $N$ antennas, and $M\gg K$. The electromagnetic waves radiated by the BS propagate in space, and the power will change with the propagation distance and environmental factors. Eventually, the power arriving at the receiver of the $k$-th UE is a physical quantity closely related to the transmit power of the BS and the channel characteristics, and can be expressed as

\begin{equation}
\mathbf{P}_{\text{u},k}^{\text{r}} = \mathbf{P}_{\text{b},k}^{\text{t}} \left| \mathbf{H}_{\text{DL},k} \right|^2,
\end{equation}
where $\mathbf{P}_{\text{b},k}^\text{t}=[p_{\text{b},1,k}^\text{t},p_{\text{b},2,k}^\text{t},\dots,p_{\text{b},M,k}^\text{t}]\in\mathbb{C}^{1\times M}$ is the matrix composed of the initial radiated electromagnetic wave power of each element of the BS, and $\mathbf{P}_{\text{u},k}^\text{r}\in\mathbb{C}^{1\times N}$ is the power matrix received by the array of the $k$-th UE. It should be emphasized that, in order to ensure the power circulation of the resonant system, the RDA of the UE will radiate part of the power, and adjust the remaining power according to the power harvesting ratio $\alpha$, the loaded power is $\mathbf{P}_{\text{u},k}^{\text{load}}=\alpha\mathbf{P}_{\text{u},k}^\text{r}$. The $\mathbf{H}_{\text{DL},k}\in\mathbb{C}^{M\times N}$ represents the MIMO channel matrix from the $M$ transmitting antennas of BS to the $N$ receiving antennas of the $k$-th UE on the sub-band $k$ \cite{zeng2017communications}. Each element of the matrix $h_{\text{DL},k}^{mn}$ can be expressed as

\begin{equation}
h_{\text{DL},k}^{mn} = \frac{\lambda_{\text{DL},k}}{4\pi} \sqrt{ G_{\text{u},n,k} G_{\text{b},m,k} r_{mn,k}^{-\beta} } e^{j \Delta \varphi_{mn,k}},
\end{equation}
where $\lambda_{\text{DL},k}$ is the electromagnetic wavelength of the $k$-th sub-band, $G_{\text{u},n,k}$ and $G_{\text{b},m,k}$ are the antenna gains of the $k$-th UE and BS, respectively. The $\beta$ represents the path loss exponent, and $\varphi_{mn}$ represents the phase shift caused by the transmission distance.
Therefore, the power of the $k$-th UE received can be expressed as
\begin{equation}
    \sum_{n = 1}^{N}[\mathbf{P}_{\text{u},k}^\text{r}]_n = \sum_{n = 1}^{N}\sum_{m = 1}^{M} p^\text{t}_{\text{b},m,k} \left(\frac{\lambda_{\text{DL},k}}{4\pi}\right)^2 G_{\text{u},n,k}G_{\text{b},m,k}r_{mn,k}^{-\beta}.
\end{equation}

According to the Institute of Radio Engineers (IRE) standards on antennas \cite{tai1961definition}, the power density matrix of the electromagnetic wave radiated from the BS to the $k$-th UE can be expressed as
\begin{equation}
\mathbf{S}_{\text{u},k} = \mathbf{P}_{\text{u},k}^{\text{r}} \oslash \mathbf{A}_k,
\label{e13}
\end{equation}
where $\mathbf{A}_k=[a_{\text{u},1,k},a_{\text{u},2,k},\dots,a_{\text{u},N,k}]\in \mathbb{C}^{1 \times N}$ represents the effective aperture area matrix. 
Notably, similar to traditional antenna design, the effective aperture of the RDA is significantly affected by its radiating elements' gain characteristics \cite{yu2010aperture}. In free space propagation, each matrix element $a_{\text{u},n,k}$ can be defined as

\begin{equation}
a_{\text{u},n,k} = \frac{G_{\text{u},n,k} \lambda_{\text{DL},k}^2}{4\pi}.
\end{equation}

Furthermore, based on the Poynting's theorem, we can obtain the electric field strength at the $k$-th UE location as follows

\begin{equation}
\mathbf{E}_{\text{u},k} = \sqrt{2 Z_0 \mathbf{S}_{\text{u},k}},
\label{e15}
\end{equation}
where $Z_0$ is the characteristic impedance, defined as the ratio of the magnitude of the electric field to the magnetic field, with a value of $377\ \Omega$ in free space \cite{ma2002electromagnetic}, we assume that all units share the same impedance. Except for the portion of power collected as energy, the remaining power is entirely used for the UL signal transmission. Thus, the power matrix of the signal transmitted by the UE is given by

\begin{equation}
\mathbf{P}_{\text{u},k}^{\text{t}} = (1 - \alpha) \mathbf{P}_{\text{b},k}^{\text{t}} \left| \mathbf{H}_{\text{DL},k} \right|^2.
\end{equation}

Owing to the inherent phase conjugation property exhibited by the RDA, the signal power transmitted by the $k$-th UE and subsequently received at the BS can be mathematically expressed as

\begin{equation}
\mathbf{P}_{\text{b},k}^{\text{r}} = \mathbf{P}_{\text{u},k}^{\text{t}} \left| \mathbf{H}_{\text{UL},k} \right|^2,
\end{equation}
where $\mathbf{H}_{\text{UL},k} \in \mathbb{C}^{N \times M}$ represents the UL channel matrix between the UE array and the BS array, with each element can be expressed as
\begin{equation}
h_{\text{UL},k}^{mn} = \frac{\lambda_{\text{UL},k}}{4\pi} \sqrt{ G_{\text{u},n,k} G_{\text{b},m,k} r_{mn,k}^{-\beta} } e^{j \Delta \varphi_{mn,k}},
\end{equation}
where $\lambda_{\text{UL},k}$ is the UL signal wavelength, and $\varphi_{mn,k}$ represents the phase shift caused by the transmission distance and field attenuation. Therefore, the power of the BS received from $K$ UEs can be expressed as

\begin{equation}
    \sum_{m = 1}^{M}[\mathbf{P}_{\text{b}}^\text{r}]_m = \sum_{k = 1}^{K}\sum_{n = 1}^{N}\sum_{m = 1}^{M}p^\text{t}_{\text{u},n,k} \left(\frac{\lambda_{\text{DL},k}}{4\pi}\right)^2 G_{\text{u},n,k}G_{\text{b},m,k}r_{mn,k}^{-\beta}.
\end{equation}

Following the same formulation procedure as in  (\ref{e13})$\textendash$(\ref{e15}), the BS electric field intensity may be characterized as

\begin{equation}
\mathbf{E}_\text{b}=\sum_{k=1}^{K}\sqrt{2Z_0\mathbf{S}_{\text{b},k}},
\end{equation}
where $\mathbf{S}_{\text{b},k}$ is the power density at which the UE's feedback signal reaches the BS location.

After receiving the signal, the BS adjusts the beamforming direction and re-transmits the power to the UE, with the transmitted power matrix expressed as

\begin{equation}
\mathbf{P}_{\text{b},k}^{\text{t}} = G_\text{PA} \mathbf{P}_{\text{b},k}^{\text{r}},
\end{equation}
where $G_\text{PA}$ is the power gain of the power amplifier equipped on the BS. Furthermore, 
the DL and UL total power transmission efficiency can be expressed as

\begin{equation}
\eta_\text{DL} =\frac{\sum_{k=1}^{K}\sum_{n=1}^N[\mathbf{P}_{\text{u},k}^{\text{r}}]_n}{\sum_{m=1}^M[\mathbf{P}_{\text{b}}^{\text{t}}]_m}    
\end{equation}

\begin{equation}
\eta_\text{UL} =\frac{\sum_{m=1}^M[\mathbf{P}_{\text{b}}^{\text{r}}]_m}{\sum_{k=1}^{K}\sum_{n=1}^N[\mathbf{P}_{\text{u},k}^{\text{t}}]_n}    
\end{equation}

\subsection{Wireless information transmission model}

In MU-RBS, the signal is initially sent by the BS, and the RF signal received by the $k$-th UE can be represented as
\begin{equation}
\mathbf{S}_{\text{u},k}^{\text{r}} = \mathbf{S}_{\text{b},k}^{\text{t}} \mathbf{H}_{\text{DL},k} + \mathbf{N}_{\text{u},k},
\label{eq15}
\end{equation}
where the matrix $\mathbf{S}^\text{t}_{\text{b},k}=[s_{\text{b},1,k}^\text{t},s_{\text{b},2,k}^\text{t},\dots,s_{\text{b},M,k}^\text{t}] \in \mathbb{C}^{1 \times M}$ represents the transmitted signal from the BS to the $k$-th UE, where each element is obtained by upconverting the baseband signal using the DL carrier frequency, i.e.
\begin{equation}
s_{\text{b},m,k}^{\text{t}}(t) = s_{\text{b},m,0}^{\text{t}}(t)e^{j 2\pi f_{\text{DL},k}t},
\end{equation}
where $f_{\text{DL},k}$ denotes the carrier frequency transmitted to the $k$-th UE. $s_{\text{b},m,0}^{\text{t}}$ represents the baseband signal transmitted by the $m$-th antenna of the BS, which can be expressed as

\begin{equation}
s_{\text{b},m,0}^{\text{t}}(t) = \sqrt{2Z_0 p_{\text{b},m,k}^{\text{t}}} e^{j \varphi_m(t)}.
\end{equation}
where the $\varphi_m(t)$ is the time-varying phase of the baseband signal transmitted by the $m$-th antenna of the BS.
 
The $\mathbf{N}_{\text{u},k} \in \mathbb{C}^{1 \times N}$ in (\ref{eq15}) represents complex Gaussian additive white noise (AWGN), and follows the distribution $\mathbf{N}_{\text{u},k} \sim \mathcal{CN}(0, \sigma_{\text{DL},k}^2 \mathbf{I})$, where the noise variance is given by

\begin{equation}
\sigma_{\text{DL},k}^2 = 2Z_0 \kappa T BF_n
\end{equation}
where $\kappa$ is the Boltzmann constant, $T$ is the ambient temperature, and $B$ is the bandwidth, $F_n$ is the noise figure of the demodulation circuit. Thus, the signal-to-noise ratio (SNR) in the DL can be expressed as

\begin{equation}
\text{SNR}_{\text{DL},k} =\sum_{n=1}^N \frac{ (1-\alpha) [P_{\text{u},k}^{\text{r}}]_n}{2Z_0 \kappa T B N},
\end{equation}

According to Shannon's capacity formula, the DL spectral efficiency can be expressed as

\begin{equation}
R_{\text{DL},k} = \log_2 \left\{1 + 10^{0.1(\text{SNR}_{\text{DL},k} - \delta)}\right\}.
\end{equation}
where $\delta$ represents the path loss factor. Similarly, the received signal at the BS for UL communication, the corresponding SNR, and spectral efficiency can be expressed by  
\begin{equation}
\text{SNR}_{\text{UL},k} = \sum_{m=1}^M\frac{ \gamma [P_{\text{b},k}^{\text{r}}]_m}{2Z_0 \kappa T B M}
\end{equation}

\begin{equation}
R_{\text{UL},k} = \log_2 \left\{1 + 10^{0.1(\text{SNR}_{\text{UL},k} - \delta)}\right\}
\end{equation}
where the $\gamma$ is the proportion of signals received by BS for communication, and the rest amplified by power amplifiers before being transmitted back.

\begin{figure*}[ht]
  \centering
          \subfigure[]{
        \includegraphics[width=0.2\linewidth]{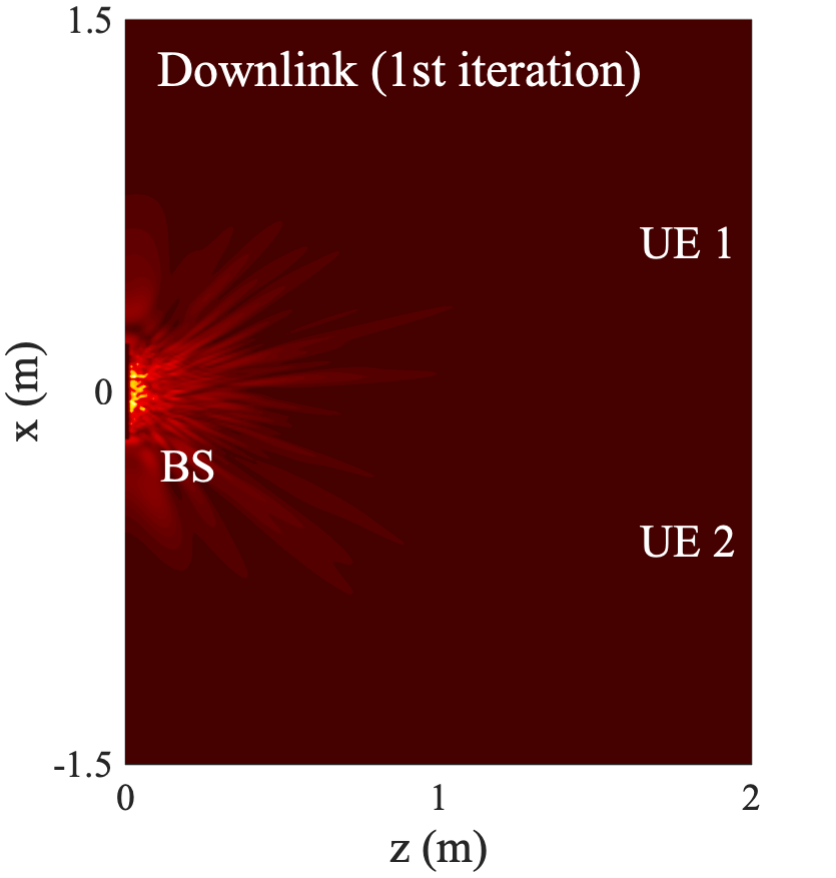}}
          \subfigure[]{
        \includegraphics[width=0.2\linewidth]{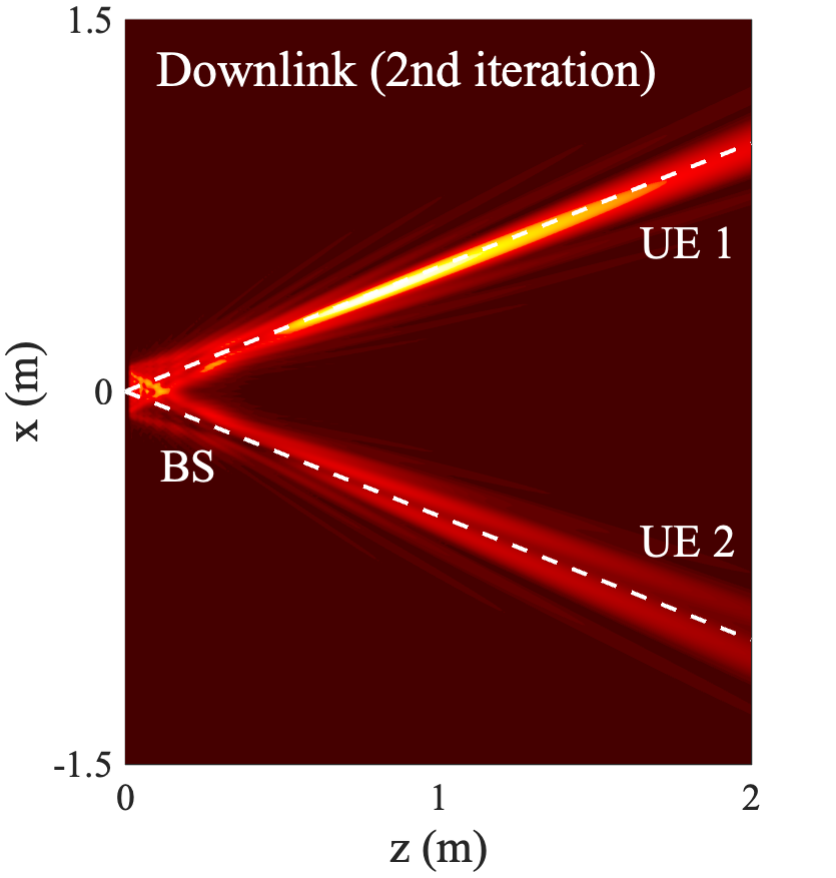}}
          \subfigure[]{
        \includegraphics[width=0.2\linewidth]{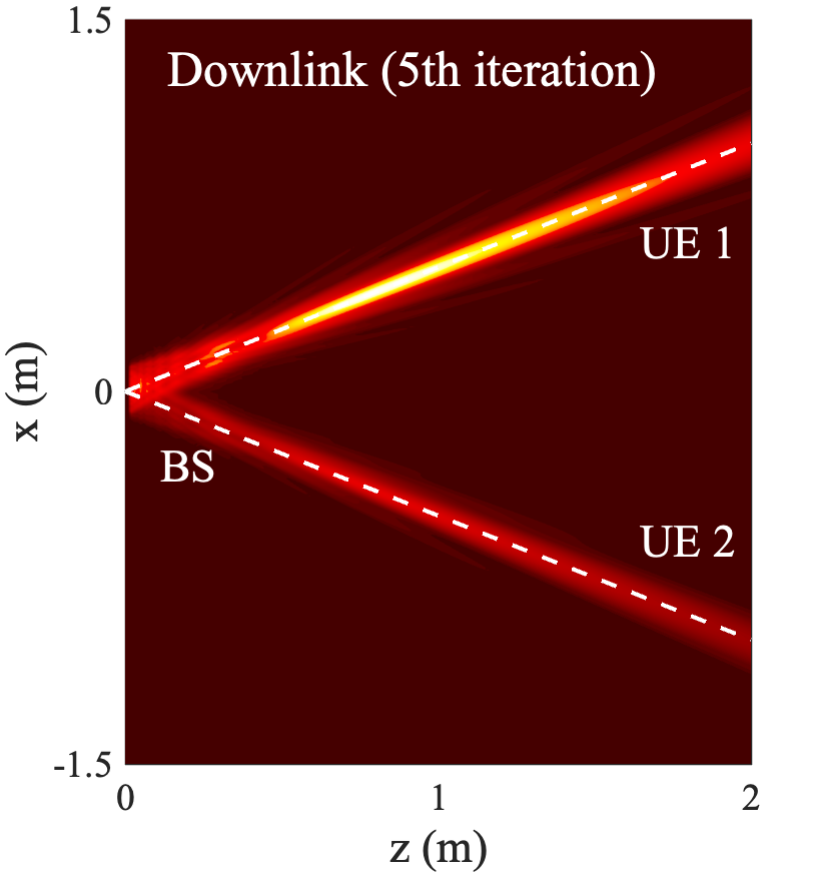}}
         \subfigure[]{
        \includegraphics[width=0.2\linewidth]{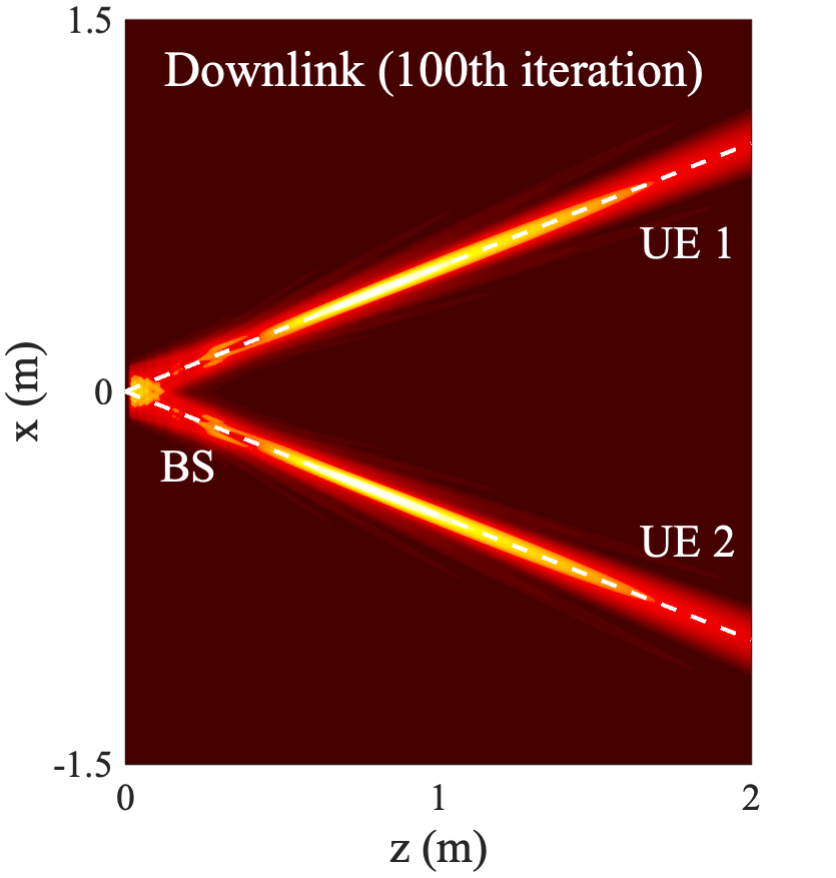}}

         \subfigure[]{
        \includegraphics[width=0.2\linewidth]{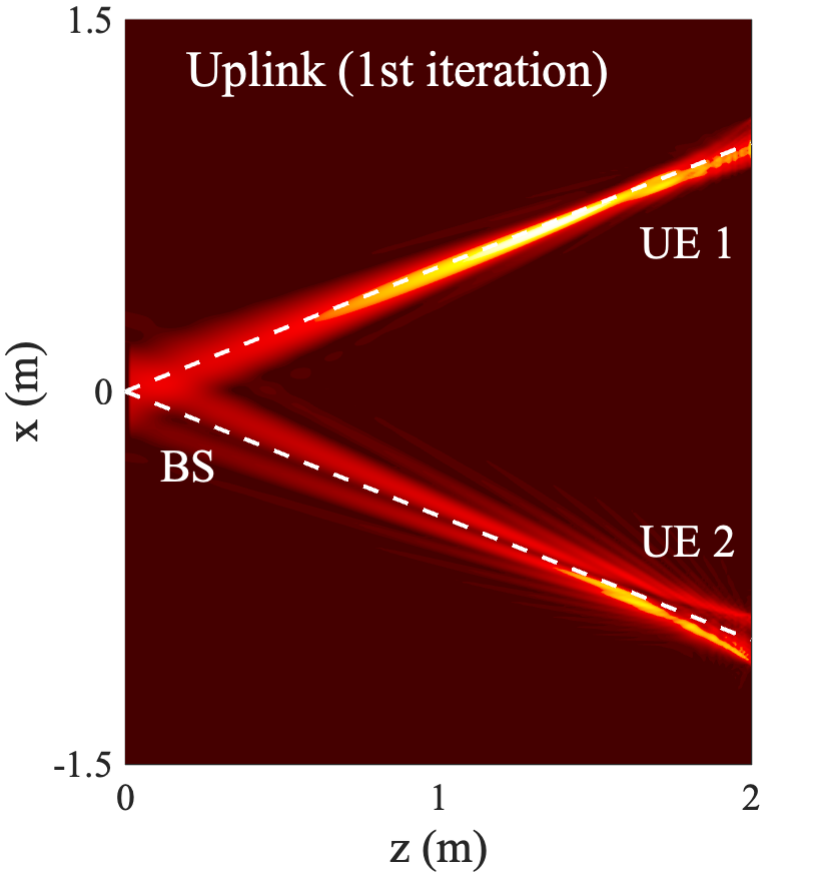}}
          \subfigure[]{
        \includegraphics[width=0.2\linewidth]{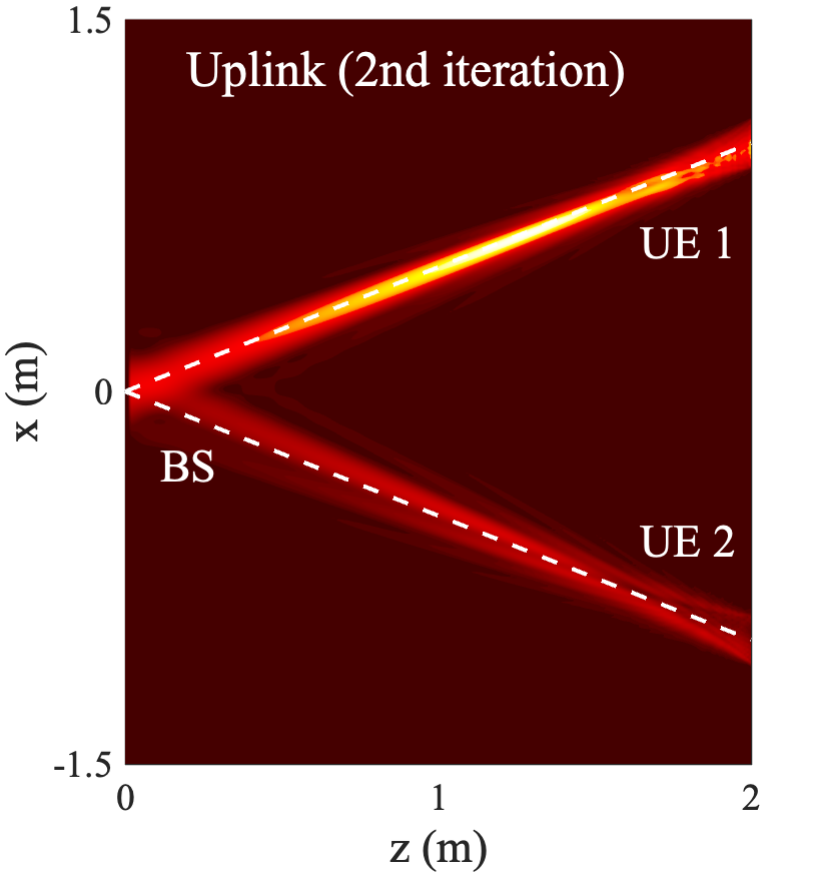}}
          \subfigure[]{
        \includegraphics[width=0.2\linewidth]{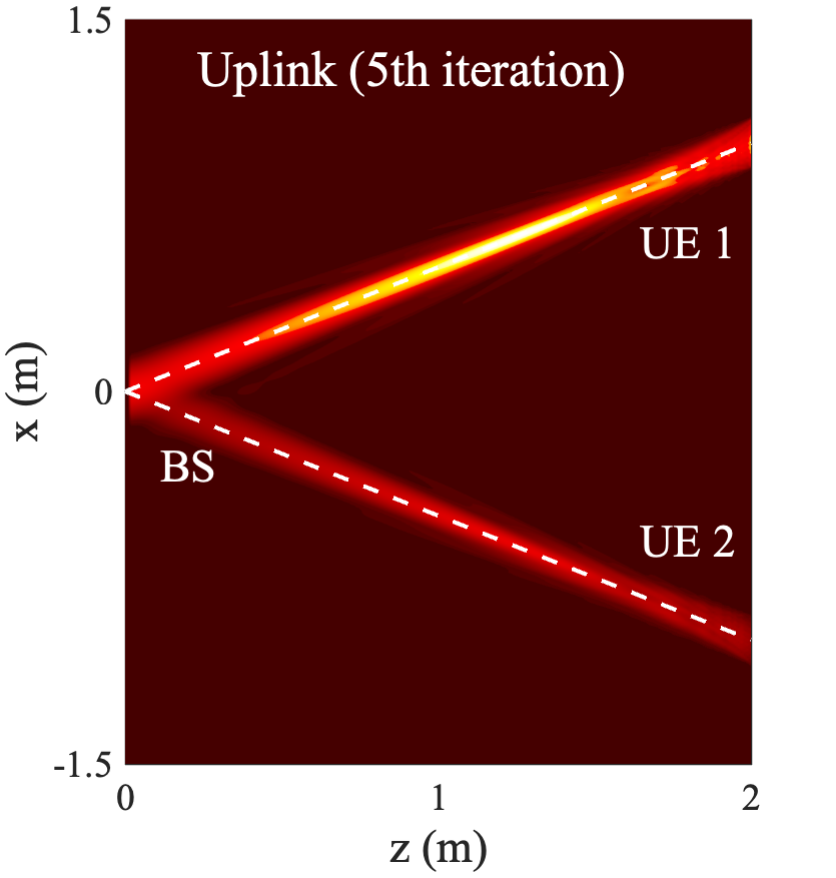}}
         \subfigure[]{
        \includegraphics[width=0.2\linewidth]{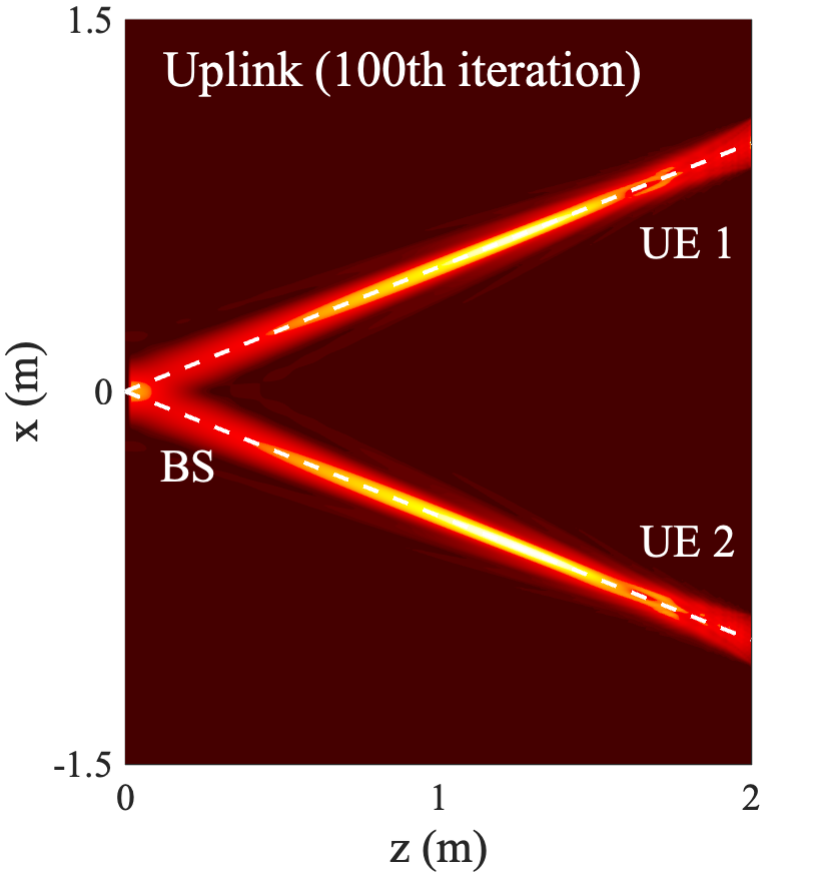}}
          \caption{Normalized spatial power density distributions in the xoz plane during SWIPT between the BS and two UEs at 1, 2, 5, and 100 iterations. Figures (a)$\textendash$(d) show the downlink, (e)$\textendash$(h) show the uplink.}  
  \label{xoz}
\end{figure*}

\section{PERFORMANCE EVALUATION}
\label{sec4}

In this section, we present MATLAB-based simulations to validate the multi-link resonance characteristics of the MU-RBS, with particular focus on analyzing the steady-state formation process and conducting comprehensive performance evaluations of both WPT and WIT. 
In the simulations, all UEs and BS array configurations follow a unified design specification: the element spacing of the Rx array and Tx array is independently set according to the half-wavelength ($\lambda_\text{UL,2}/2$, $\lambda_\text{DL,2}/2$) of the UL and DL operating frequencies for UE 2, respectively. Since the Tx and Rx element spacing ratios are mismatched, no element overlap occurs in our simulations. All simulation parameters are configured as specified in TABLE~\ref{tab:parameter_setting} (unless otherwise noted).

The array configurations for all UEs and BS follow a unified design specification: the element spacing of the Rx and Tx arrays are independently set to half-wavelength ($\lambda_\text{UL,2}/2$ and $\lambda_\text{DL,2}/2$, respectively) based on the uplink and downlink operating frequencies of UE 2.
For UE 1, we maintain the same element spacing configuration while deriving new UL and DL operating frequencies according to the wavelength-spacing relationship shown in (\ref{77}), i.e. ${d_\text{rx}}/{d_\text{tx}} ={\lambda_\text{in}}/{\lambda_\text{out}}$.
Since the Tx and Rx element spacings are proportionally mismatched, no element overlap occurs in our simulations.

\begin{table}
    \centering
    \caption{Parameter Setting}
    \begin{tabular}{m{3.5cm}<{\centering} m{1.5cm}<{\centering} m{2cm}<{\centering}}
        \toprule 
        \textbf{Parameter} & \textbf{Symbol} & \textbf{Value} \\
        \midrule
        Antenna gain& $G$ & $\leq$4.97 dBi \\     
        UE 1 downlink frequency & $f_{\text{DL},1}$ &  28.517 GHz\\
        UE 1 uplink frequency & $f_{\text{UL},1}$ &  29.5 GHz\\
        UE 2 downlink frequency & $f_{\text{DL},2}$ &  29 GHz\\
        UE 2 uplink frequency & $f_{\text{UL},2}$ &  30 GHz\\    
        Rx element spacing & $\lambda_\text{UL,2}/2$ & 0.50 cm \\
        Tx element spacing & $\lambda_\text{DL,2}/2$ & 0.52 cm \\     
        Power harvesting ratio & $\alpha$ & 99.5\% \\
        Signal processing ratio & $\gamma$ & 99.5\% \\
        UEs' elevation & $\theta_1$, $\theta_2$ & $10^\circ$, $-10^\circ$\\
        Path loss exponent & $\beta$ & 2 \\
        Ambient temperature  & $T$ & 295 K\\
        Path loss factor  & $\delta$ & 3 dB \\
        Bandwidth  & $B$ & 1 GHz \\
        Demodulation noise figure  & $F_n$ & 6 dB \\
        Boltzmann constant &$\kappa$& $1.38\times 10^{-23}$\\
        \bottomrule
    \end{tabular}
    \label{tab:parameter_setting}
\end{table}

\begin{figure*}[ht]
  \centering
        \subfigure[]{
        \includegraphics[width=0.2\linewidth]{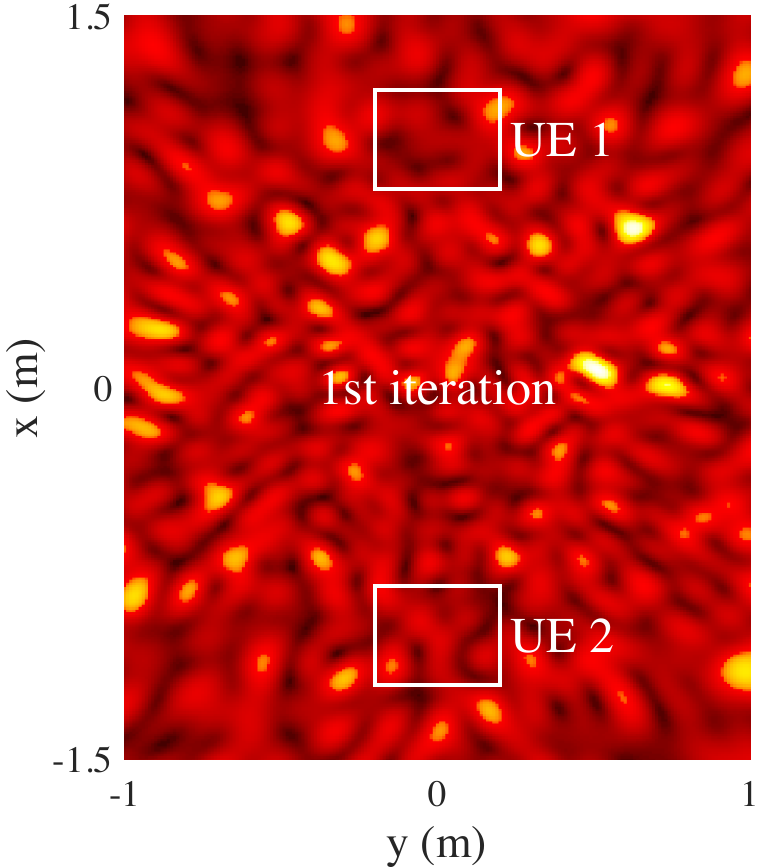}}
          \subfigure[]{
        \includegraphics[width=0.2\linewidth]{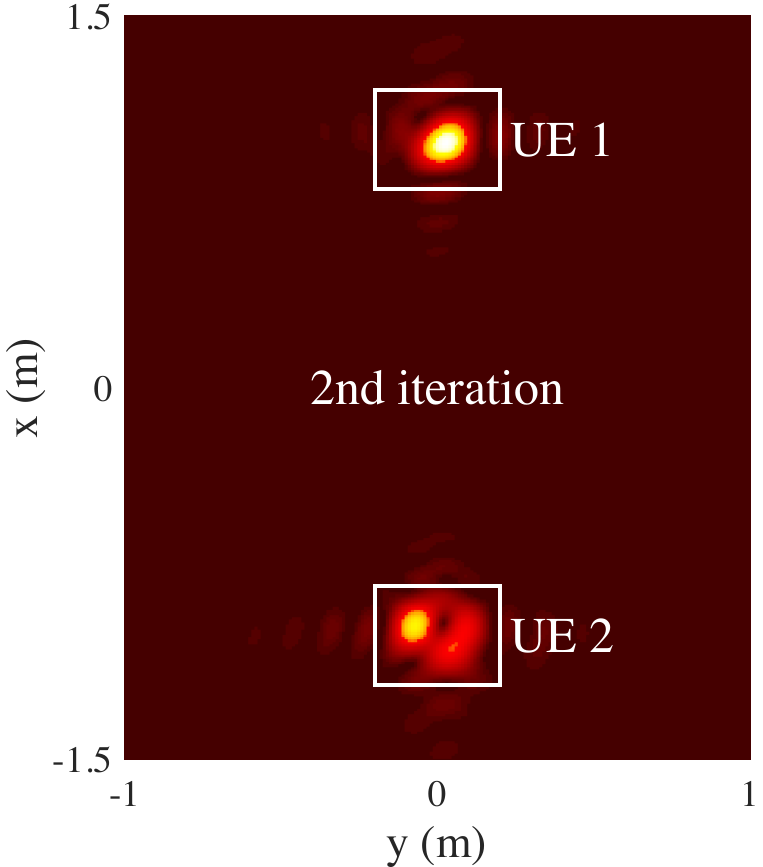}}
          \subfigure[]{
        \includegraphics[width=0.2\linewidth]{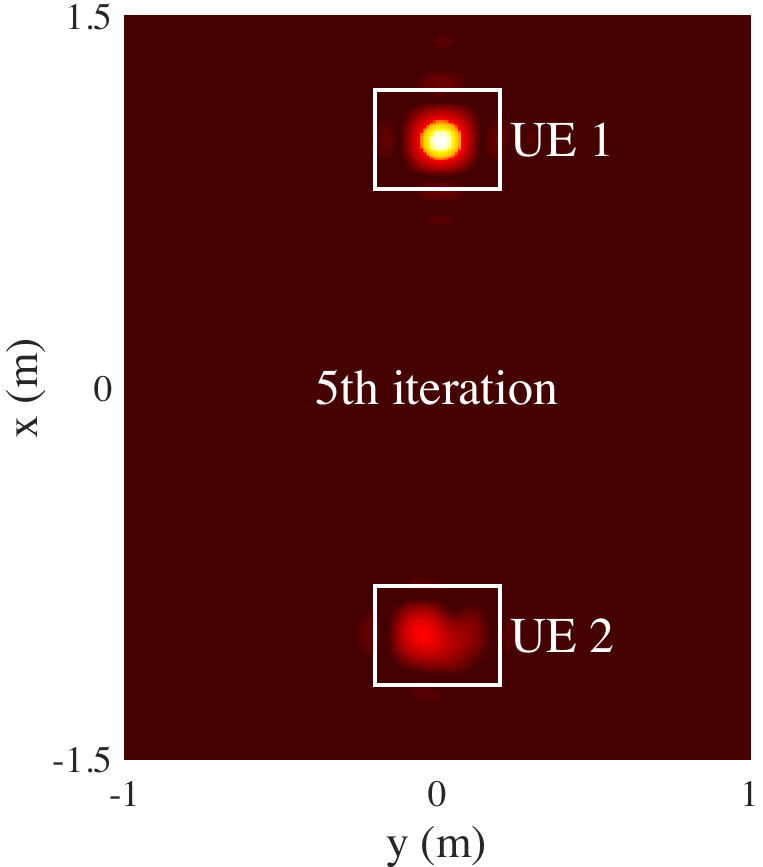}}
         \subfigure[]{
        \includegraphics[width=0.2\linewidth]{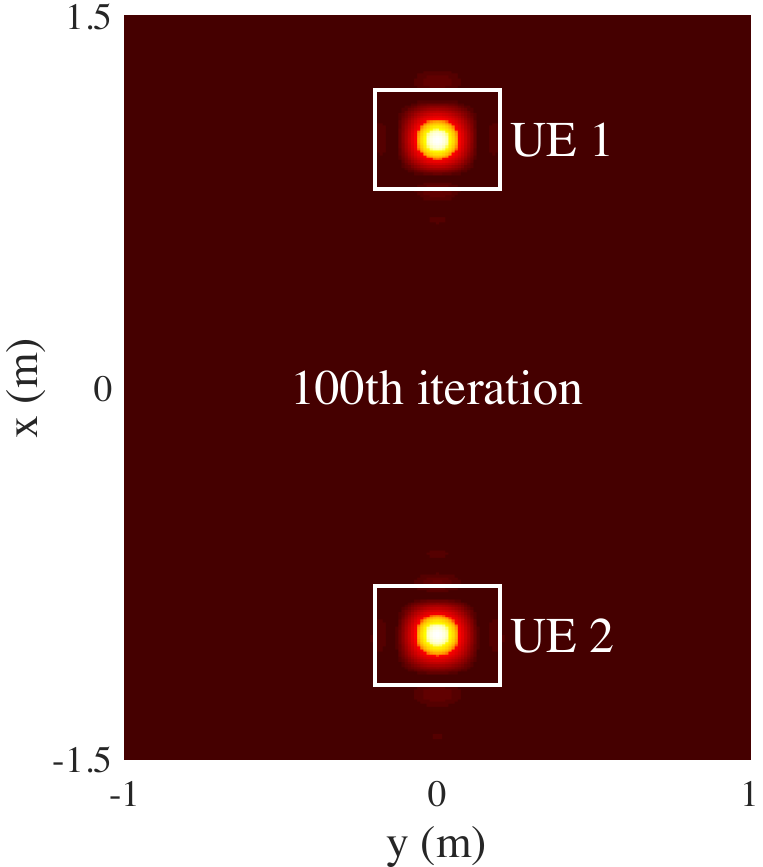}}
        
        \subfigure[]{
        \includegraphics[width=0.2\linewidth]{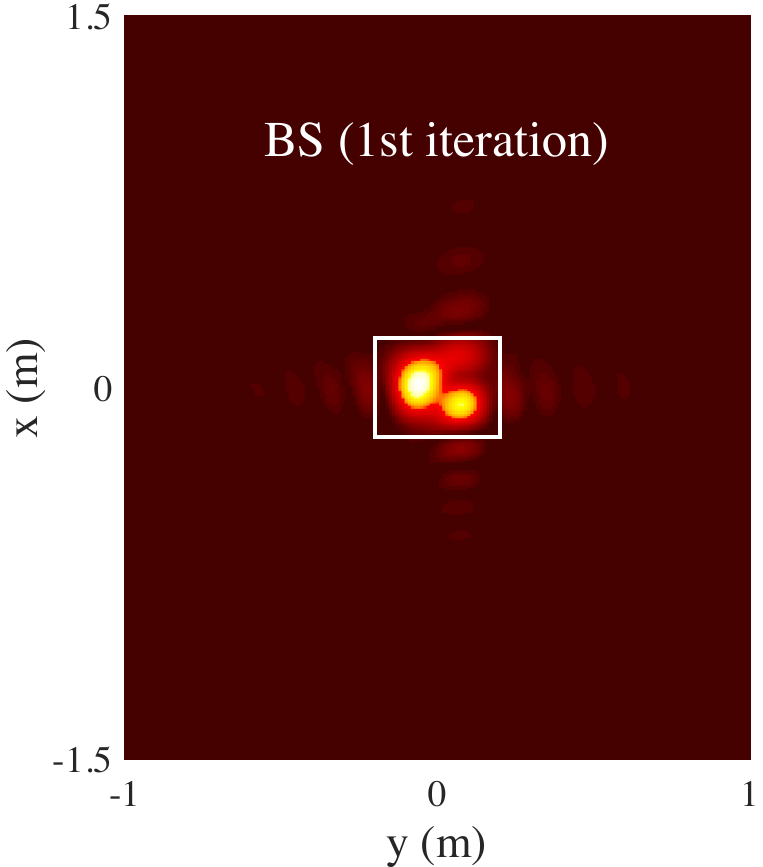}}
          \subfigure[]{
        \includegraphics[width=0.2\linewidth]{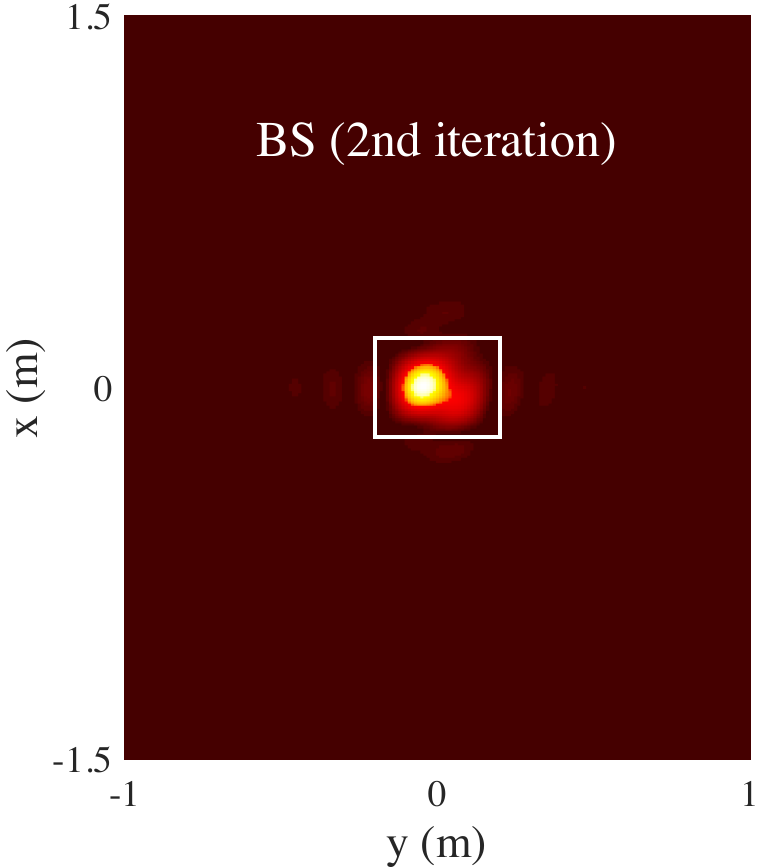}}
          \subfigure[]{
        \includegraphics[width=0.2\linewidth]{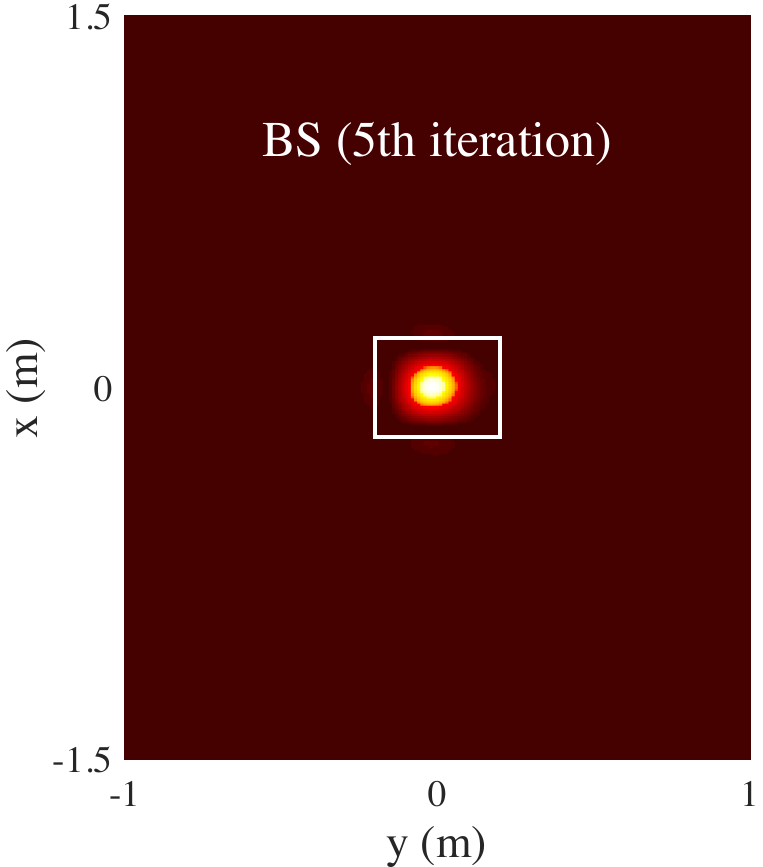}}
         \subfigure[]{
        \includegraphics[width=0.2\linewidth]{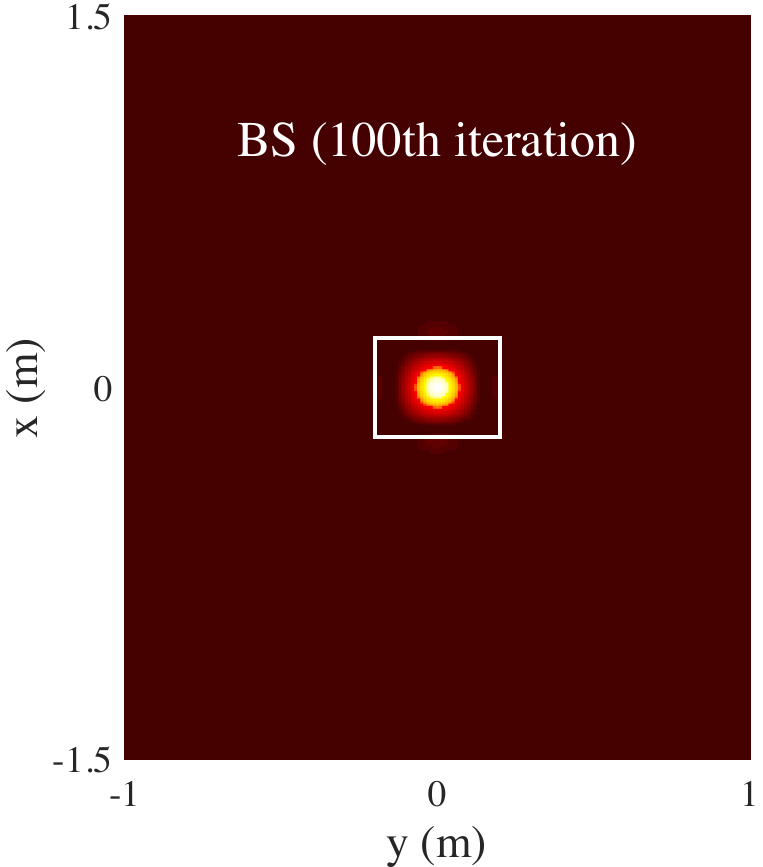}} 
          \caption{Normalized spatial power density distributions in the BS plane and UE plane during SWIPT at 1, 2, 5, and 100 iterations. Figures (a)$\textendash$(d) show the UE plane, (e)$\textendash$(h) show the BS plane.}  
  \label{xoy}
\end{figure*}
Figure \ref{xoz} illustrates the normalized spatial power density distribution in the xoz plane during wireless information and energy transmission between the BS and two UEs. The BS is located at the origin, while UE 1 and UE 2 are positioned at coordinates (1, 2, 0) and (-1, 2, 0), respectively. Figure \ref{xoz} (a)$\textendash$(d) depict the power density distribution of the DL. Initially, the signal is emitted omnidirectionally from the BS. As the iterations progress, the beams gradually align with the UEs, and the power density distribution becomes more uniform. Figure \ref{xoz} (e)$\textendash$(h) show the power density distribution during the UL transmission after the UEs receive the signal radiated from the BS. Without actively transmitting signals, the UEs demonstrate initial beam steering capability in the first iteration. As the iterations continue, the sidelobes of the beams gradually diminish, and the power density distribution becomes more uniform.

To further demonstrate the beam alignment performance of the MU-RBS during the iteration process, Figure~\ref{xoy} presents the normalized power density distribution in the xoy plane corresponding to Figure~\ref{xoz}. Figure~\ref{xoy} (a)$\textendash$(d) correspond to the UE plane (z = 2), while Figure~\ref{xoy} (e)$\textendash$(f) correspond to the BS plane. The white squares indicate the array positions. In Figure~\ref{xoy} (a), the electromagnetic waves initially radiated omnidirectionally from the BS appear scattered in the UE plane without clear directionality. After the UEs receive the signal and reflect it back to the BS, the power density distribution in the BS plane, as shown in Figure~\ref{xoy} (e), exhibits weak beam directionality and numerous sidelobes due to the random phases of the signals received by the UEs. However, as the iterations progress, it becomes evident that the beams in both the UE and BS planes focus on the centers of the UEs and BS, respectively, with significantly reduced sidelobes.

\begin{figure}
    \centering
    \includegraphics[width=\linewidth]{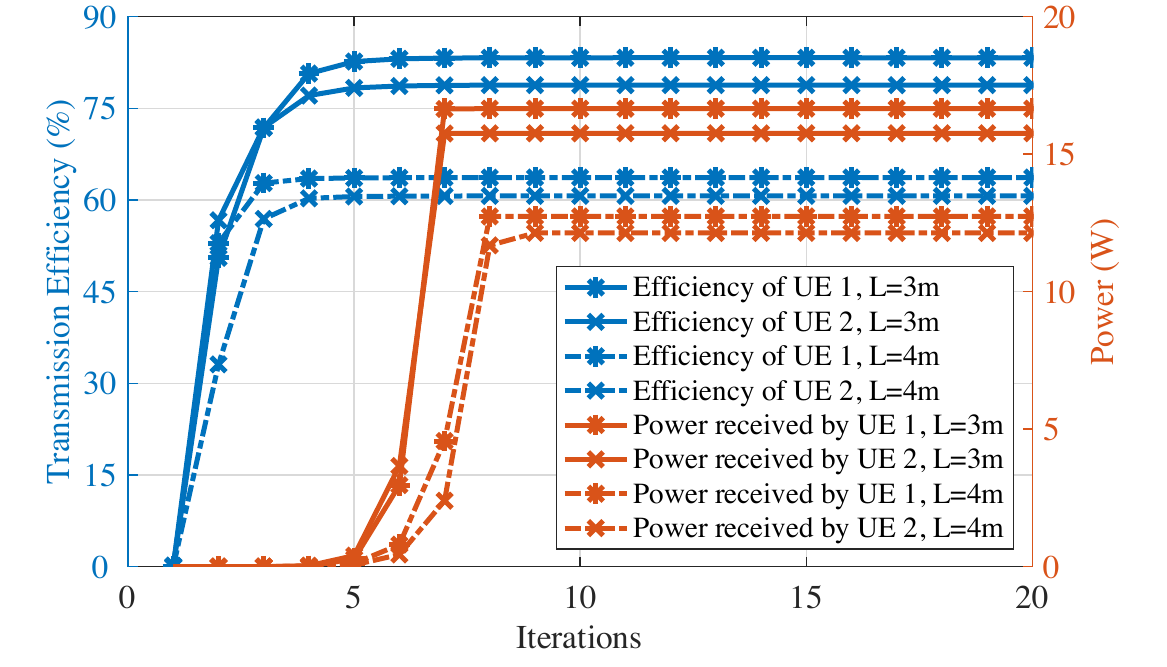}
    \caption{Performance of WPT at different distances during the iteration process.}
    \label{Iteration_power}
\end{figure}

\begin{figure*}[ht]
  \centering
  \subfigure[]{
        \includegraphics[width=0.9\linewidth]{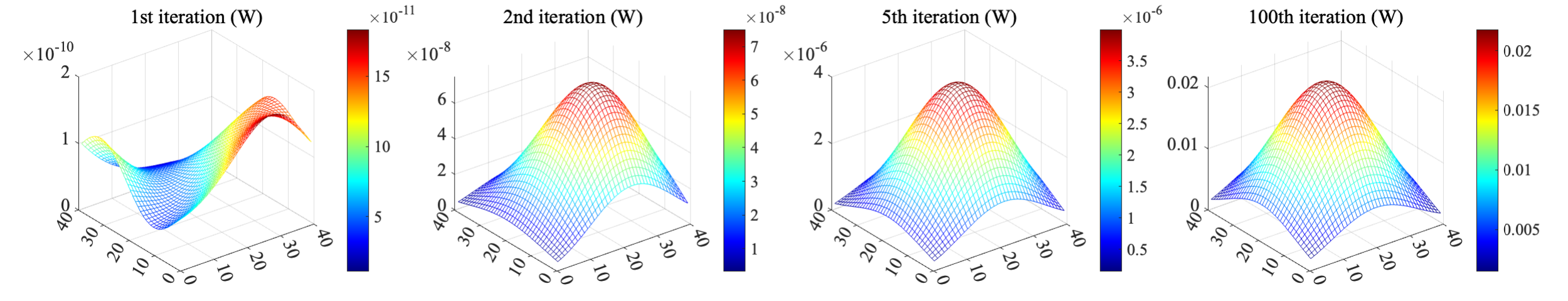}}
          \subfigure[]{
        \includegraphics[width=0.9\linewidth]{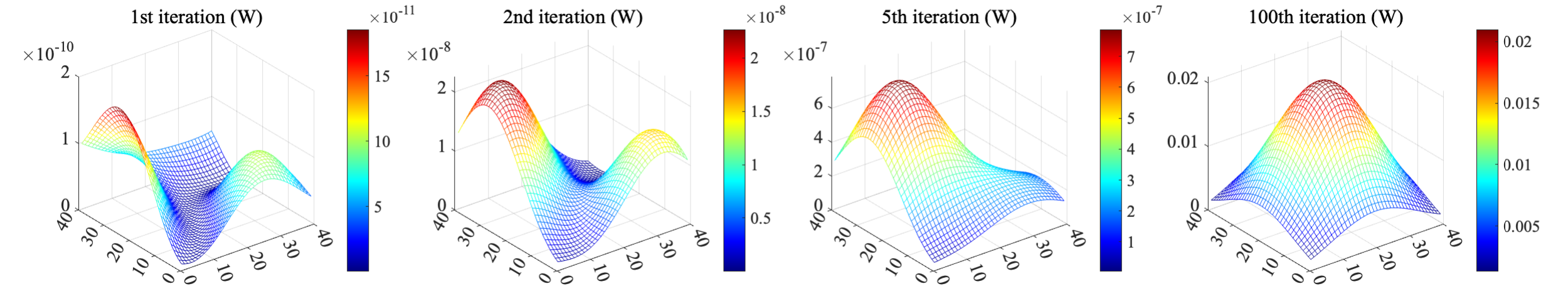}}
  \caption{The received signal power distributions of UEs during the iterative process. (a) Signal power distribution of UE 1; (b) signal power distribution of UE 2.
}  
  \label{UEPower}
\end{figure*}

Figure \ref{Iteration_power} presents the WPT
performance at different distances during the iteration process. The left side illustrates the power transmission efficiency from the BS to each UE. It can be observed that the transmission efficiency increases rapidly in the initial iterations and eventually reaches a steady state. The transmission efficiency of UE 1 consistently outperforms that of UE 2, which is attributed to the reduction in channel gain caused by higher frequencies. Similarly, distance also leads to a decline in transmission efficiency, with the efficiency dropping to approximately 60\% at 4 m. The right side shows the received power at each UE during the iteration process. By comparison, it can be seen that the number of iterations required for the received power of each UE to stabilize is higher than that for the transmission efficiency. This is because, although the round-trip beam phases align quickly through iterations, the BS's gain has not yet compensated for the losses, and the amplifier continues to operate at high gain until the amplifier gain $G_\text{PA}$ decreases to 0 dB, at which point the system achieves the optimal energy transfer state.

Furthermore, the reduction in transmission efficiency caused by longer distances and higher frequencies slows the growth rate of the input power at the UEs. This gap becomes more pronounced as the iterations progress. It is important to note that a maximum output power limit has been imposed to ensure the safety of WPT, such that the output power of the BS does not exceed 20 W.

Figure \ref{UEPower} illustrates the signal power distributions of UE~1 and UE~2 (depicted in (a) and (b), respectively) during the iterative process. UE~1 demonstrates rapid convergence of its power distribution towards the array center, with power steadily increasing to reach a peak as the iteration advances. Conversely, UE~2 requires a greater number of iterations to achieve a similar ``focusing'' of its power distribution. This disparity indicates that the system exhibits a more pronounced optimization effect for low frequency UEs compared to high frequency UEs.

\begin{figure}[!t]
  \centering
        \subfigure[]{
        \includegraphics[width=\linewidth]{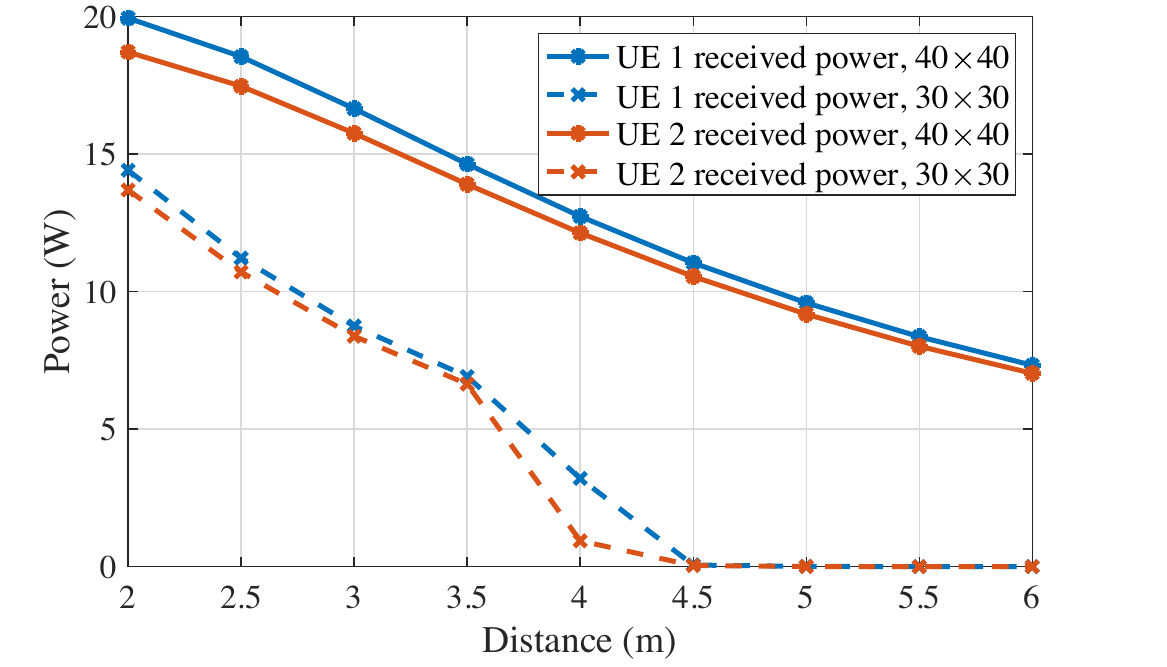}}
         \subfigure[]{
        \includegraphics[width=\linewidth]{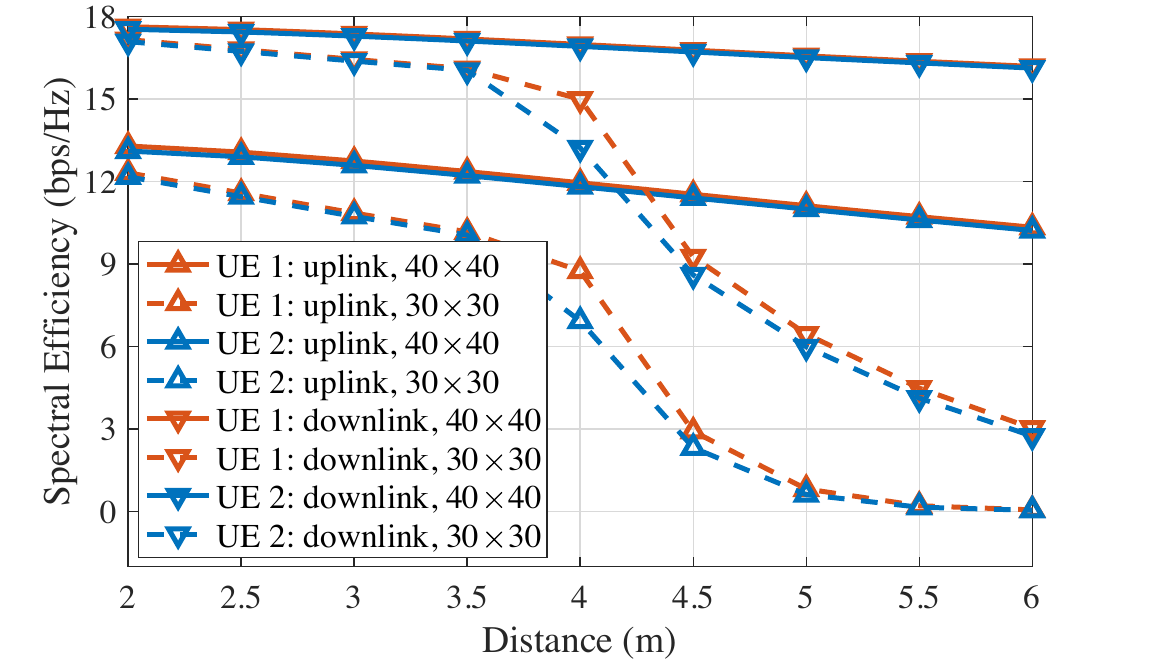}}  
  \caption{Variations in WPT and WIT performance of MU-RBS with different array sizes under different link lengths at steady state. (a) Power transmission efficiency versus distance; (b) received power versus distance; (c) SNR versus distance; (d) spectral efficiency versus distance.}  
  \label{Distance}
\end{figure}

Figure \ref{Distance} presents the variation of UEs received signal power and UL/DL spectral efficiency with link length for MU-RBS with different array sizes after reaching steady state.
As shown in Figure \ref{Distance}~(a), when the link length is 2 m, the UEs received power is maximized due to high transmission efficiency and the BS output power reaching the preset maximum of 20 W. For the same array size, the received power of low frequency UEs is consistently higher than that of high frequency UEs; when the link length exceeds 4.5 m, the received power of UEs with a 30$\times$30 array approaches 0 W.

Figure \ref{Distance}~(b) indicates that for the same array size, the DL spectral efficiency is always higher than the UL spectral efficiency. This is because the UE side operates in a passive reflection mode, only returning a small portion of the power of electromagnetic waves, while the BS continuously transmits high power signals. Additionally, the spectral efficiency difference between the two UEs is small at different distances, but this difference significantly increases when the array size is smaller. When the link length exceeds 4.5 m, the UL and DL spectral efficiencies of the system with a 30$\times$30 array configuration approach 0 bps/Hz, which is consistent with the attenuation trend of UEs received power.
In summary, for practical deployments, to maximize the effective transmission distance of WPT and WIT, it is recommended to increase the array specification or raise the BS maximum output power to ensure transmission efficiency. It should be noted that excessively high output power may trigger safety hazards for short range transmissions.

\begin{figure*}[ht]
  \centering
  \subfigure[]{
        \includegraphics[width=0.35\linewidth]{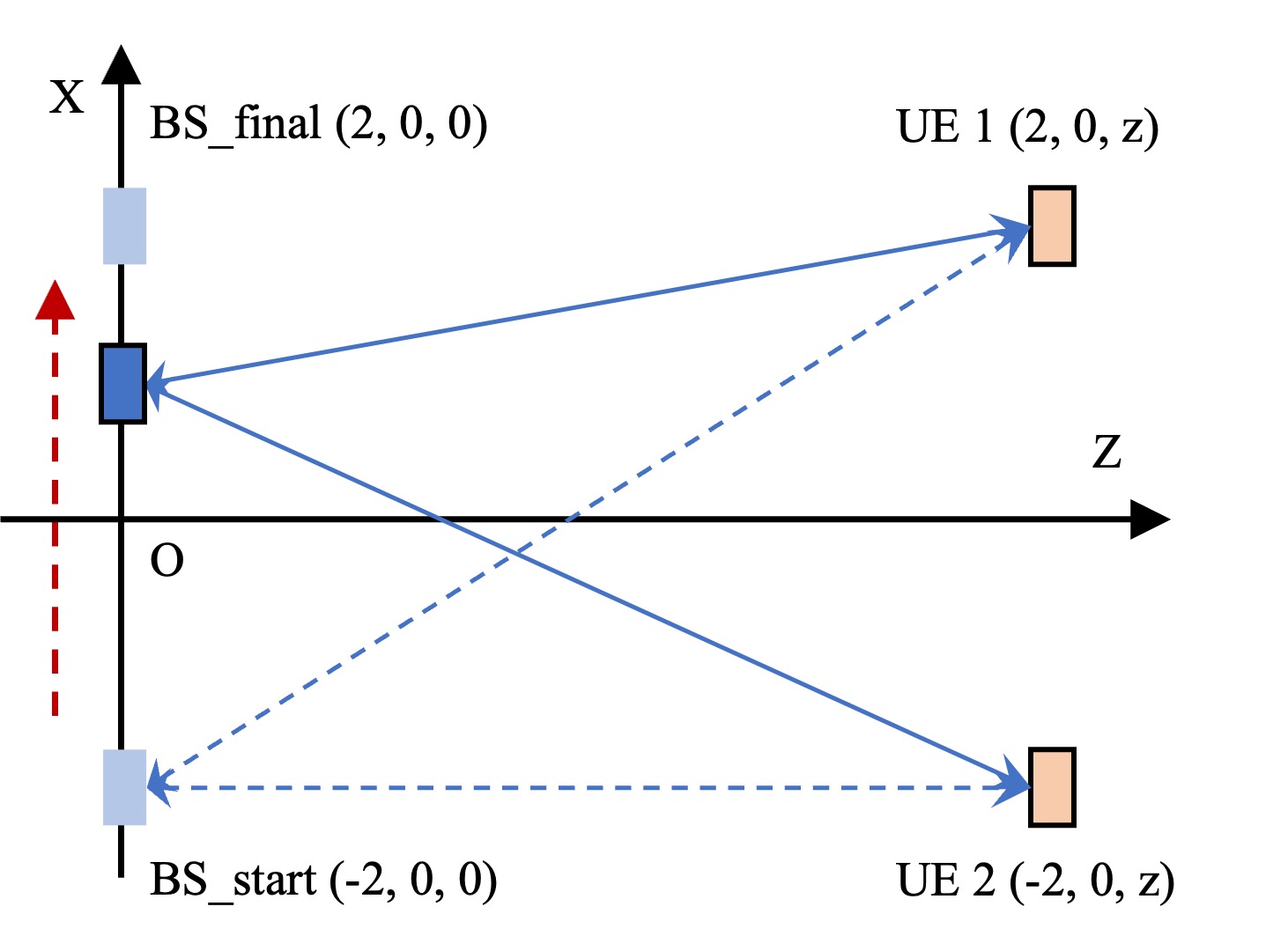}}
          \subfigure[]{
        \includegraphics[width=0.45\linewidth]{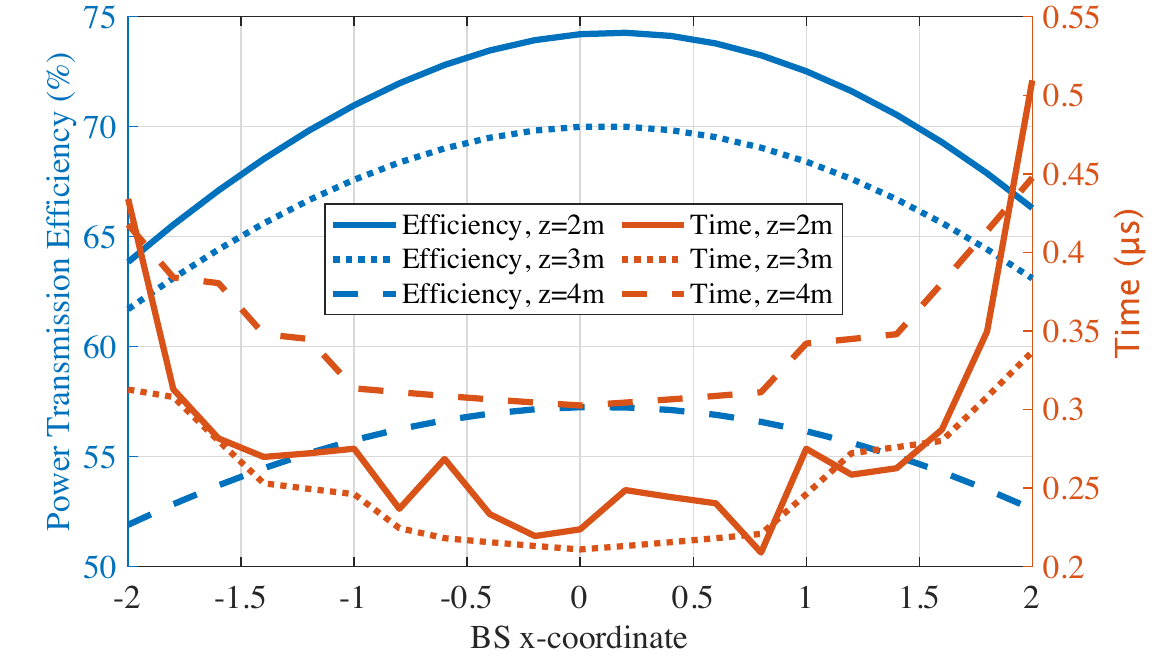}}
  \caption{Variations in system power transmission efficiency and resonance establishment time as the BS moves along the x-axis. (a) Schematic diagram of movement;
(b) power transmission efficiency and time versus the x-coordinate of the BS.}  
  \label{mobile}
\end{figure*}

In Figure~\ref{mobile}, the BS is moved along the x-axis of a three-dimensional Cartesian coordinate system, as shown in Figure~\ref{mobile} (a). It starts from the initial position BS\_star at (-2, 0, 0) and moves to BS\_final at (2, 0, 0), with coordinates (x, 0, 0). The coordinates of UE 1 and UE 2 are (2, 0, z) and (-2, 0, z), respectively. We then analyze the total power transmission efficiency of the system and the time required to establish resonance. The total system transmission efficiency is calculated as the sum of the power received by all UEs divided by the total power transmitted by the BS after resonance is established with all UEs. The time required to establish resonance depends on the transmission distance and the number of iterations, which can be expressed as
\begin{equation}
    T=\max\{(2L_1+I_1)/c,(2L_2+I_2)/c\}
\end{equation}
 where the $L_1$ and $L_2$ are the distance between the BS and UEs, $I_1$ and $I_2$ are the number of iterations corresponding to UE 1 and UE 2.

 From the Figure~\ref{mobile}, it can be observed that the total system transmission efficiency is highest when the BS is positioned midway between UE 1 and UE 2. When the BS moves closer to either UE 1 or UE 2, the efficiency decreases. However, the rate of efficiency degradation is significantly slower when the BS approaches the lower frequency UE 1 compared to the higher frequency UE 2.

 The trend in resonance establishment time with respect to BS movement is opposite to that of transmission efficiency. The time required to establish resonance is shortest when the BS is located midway between UE 1 and UE 2, as the distances from the BS to both UEs are equal, resulting in the fewest iterations for convergence. Additionally, it can be seen that the resonance establishment time is longest at $z$=4 m, but it is not the shortest at $z$=2 m. This is because, due to near-field effects, electromagnetic waves approximate plane waves more closely at longer distances, reducing the number of iterations required for phase alignment. Therefore, in practical deployments of MU-RBS, the BS should ideally be positioned midway between the two UEs or closer to the lower frequency UE. This approach can effectively enhance the system's WPT performance and reduce the time required to establish resonance.
 
\section{conclusion}
\label{sec5}
The MU-RBS proposed in the paper effectively solves the problems of complex target detection, low transmission efficiency, and co-frequency interference in traditional SWIPT systems in passive scenarios. By using RDA equipped with BS and UEs, combined with FDMA technology, multi-user adaptive beam alignment and resonance can be achieved. From the simulation results, it can be seen that the system exhibits good convergence during the iteration process, which can not only achieve efficient wireless power transmission, but also wireless information transmission with high spectral efficiency. This design is expected to provide new ideas and theoretical support for the application of next-generation wireless communication technology in passive scenarios.

In future work, we will further explore transmission schemes for NLOS channels and pursue practical design and implementation of PLL-divided conjugate antenna arrays.
\bibliographystyle{IEEEtran}

\bibliography{Mybib}

\begin{thebibliography}{10}
\providecommand{\url}[1]{#1}
\csname url@samestyle\endcsname
\providecommand{\newblock}{\relax}
\providecommand{\bibinfo}[2]{#2}
\providecommand{\BIBentrySTDinterwordspacing}{\spaceskip=0pt\relax}
\providecommand{\BIBentryALTinterwordstretchfactor}{4}
\providecommand{\BIBentryALTinterwordspacing}{\spaceskip=\fontdimen2\font plus
\BIBentryALTinterwordstretchfactor\fontdimen3\font minus \fontdimen4\font\relax}
\providecommand{\BIBforeignlanguage}[2]{{%
\expandafter\ifx\csname l@#1\endcsname\relax
\typeout{** WARNING: IEEEtran.bst: No hyphenation pattern has been}%
\typeout{** loaded for the language `#1'. Using the pattern for}%
\typeout{** the default language instead.}%
\else
\language=\csname l@#1\endcsname
\fi
#2}}
\providecommand{\BIBdecl}{\relax}
\BIBdecl

\bibitem{wikstrom2020challenges}
G.~Wikstr{\"o}m, J.~Peisa, P.~Rugeland, N.~Johansson, S.~Parkvall, M.~Girnyk, G.~Mildh, and I.~L. Da~Silva, ``Challenges and technologies for 6{G},'' in \emph{2020 2nd 6G wireless summit (6G SUMMIT)}.\hskip 1em plus 0.5em minus 0.4em\relax IEEE, 2020, pp. 1--5.

\bibitem{shinohara2020trends}
N.~Shinohara, ``Trends in wireless power transfer: {WPT} technology for energy harvesting, mllimeter-wave/thz rectennas, {MIMO-WPT}, and advances in near-field {WPT} applications,'' \emph{IEEE microwave magazine}, vol.~22, no.~1, pp. 46--59, 2020.

\bibitem{9675011}
B.~Clerckx, J.~Kim, K.~W. Choi, and D.~I. Kim, ``Foundations of wireless information and power transfer: Theory, prototypes, and experiments,'' \emph{Proceedings of the IEEE}, vol. 110, no.~1, pp. 8--30, 2022.

\bibitem{an2021energy}
H.~An and H.~Park, ``Energy-balancing resource allocation for wireless cooperative iot networks with {SWIPT},'' \emph{IEEE Internet of Things Journal}, vol.~9, no.~14, pp. 12\,258--12\,271, 2021.

\bibitem{perera2017simultaneous}
T.~D.~P. Perera, D.~N.~K. Jayakody, S.~K. Sharma, S.~Chatzinotas, and J.~Li, ``Simultaneous wireless information and power transfer ({SWIPT}): Recent advances and future challenges,'' \emph{IEEE Communications Surveys \& Tutorials}, vol.~20, no.~1, pp. 264--302, 2017.

\bibitem{yang2017magnetic}
G.~Yang, M.~R.~V. Moghadam, and R.~Zhang, ``Magnetic {MIMO} signal processing and optimization for wireless power transfer,'' \emph{IEEE Transactions on Signal Processing}, vol.~65, no.~11, pp. 2860--2874, 2017.

\bibitem{yang2018wireless}
L.~Yang, Y.~Zeng, and R.~Zhang, ``Wireless power transfer with hybrid beamforming: How many {RF} chains do we need?'' \emph{IEEE Transactions on Wireless Communications}, vol.~17, no.~10, pp. 6972--6984, 2018.

\bibitem{moghadam2018energy}
N.~N. Moghadam, G.~Fodor, M.~Bengtsson, and D.~J. Love, ``On the energy efficiency of {MIMO} hybrid beamforming for millimeter-wave systems with nonlinear power amplifiers,'' \emph{IEEE Transactions on Wireless Communications}, vol.~17, no.~11, pp. 7208--7221, 2018.

\bibitem{zhang2018wireless}
H.~Zhang, Y.-X. Guo, S.-P. Gao, and W.~Wu, ``Wireless power transfer antenna alignment using third harmonic,'' \emph{IEEE Microwave and Wireless Components Letters}, vol.~28, no.~6, pp. 536--538, 2018.

\bibitem{mitani2019experimental}
T.~Mitani, S.~Kawashima, and N.~Shinohara, ``Experimental study on a retrodirective system utilizing harmonic reradiation from rectenna,'' \emph{IEICE Transactions on Electronics}, vol. 102, no.~10, pp. 666--672, 2019.

\bibitem{zeng2017communications}
Y.~Zeng, B.~Clerckx, and R.~Zhang, ``Communications and signals design for wireless power transmission,'' \emph{IEEE Transactions on Communications}, vol.~65, no.~5, pp. 2264--2290, 2017.

\bibitem{ju2013throughput}
H.~Ju and R.~Zhang, ``Throughput maximization in wireless powered communication networks,'' \emph{IEEE Transactions on Wireless Communications}, vol.~13, no.~1, pp. 418--428, 2013.

\bibitem{lee2017retrodirective}
S.~Lee, Y.~Zeng, and R.~Zhang, ``Retrodirective multi-user wireless power transfer with massive {MIMO},'' \emph{IEEE Wireless Communications Letters}, vol.~7, no.~1, pp. 54--57, 2017.

\bibitem{wu2019towards}
Q.~Wu and R.~Zhang, ``Towards smart and reconfigurable environment: Intelligent reflecting surface aided wireless network,'' \emph{IEEE communications magazine}, vol.~58, no.~1, pp. 106--112, 2019.

\bibitem{wu2021intelligent}
Q.~Wu, S.~Zhang, B.~Zheng, C.~You, and R.~Zhang, ``Intelligent reflecting surface-aided wireless communications: A tutorial,'' \emph{IEEE transactions on communications}, vol.~69, no.~5, pp. 3313--3351, 2021.

\bibitem{gong2020toward}
S.~Gong, X.~Lu, D.~T. Hoang, D.~Niyato, L.~Shu, D.~I. Kim, and Y.-C. Liang, ``Toward smart wireless communications via intelligent reflecting surfaces: A contemporary survey,'' \emph{IEEE Communications Surveys \& Tutorials}, vol.~22, no.~4, pp. 2283--2314, 2020.

\bibitem{gao2022beamforming}
Y.~Gao, Q.~Wu, G.~Zhang, W.~Chen, D.~W.~K. Ng, and M.~Di~Renzo, ``Beamforming optimization for active intelligent reflecting surface-aided {SWIPT},'' \emph{IEEE Transactions on Wireless Communications}, vol.~22, no.~1, pp. 362--378, 2022.

\bibitem{wu2020joint}
Q.~Wu and R.~Zhang, ``Joint active and passive beamforming optimization for intelligent reflecting surface assisted {SWIPT} under {QoS} constraints,'' \emph{IEEE Journal on Selected Areas in Communications}, vol.~38, no.~8, pp. 1735--1748, 2020.

\bibitem{10924145}
W.~Fang, W.~Chen, Q.~Wu, X.~Zhu, Q.~Wu, and N.~Cheng, ``Channel characterization of {IRS}-assisted resonant beam communication systems,'' \emph{IEEE Transactions on Communications}, pp. 1--1, 2025.

\bibitem{jiang2025single}
Q.~Jiang, M.~Liu, M.~Xu, W.~Fang, M.~Xiong, Q.~Liu, and S.~Zhou, ``Single-frequency self-alignment rf resonant beam for information and power transfer,'' \emph{IEEE Internet of Things Journal}, vol.~12, no.~11, pp. 16\,622--16\,636, 2025.

\bibitem{xia2024millimeter}
S.~Xia, Q.~Jiang, W.~Fang, Q.~Liu, S.~Zhou, M.~Liu, and M.~Xiong, ``Millimeter-wave resonant beam swipt,'' \emph{IEEE Internet of Things Journal}, vol.~11, no.~24, pp. 40\,464--40\,477, 2024.

\bibitem{buchanan2011high}
N.~Buchanan, V.~Fusco, and M.~Van Der~Vorst, ``A high performance analogue retrodirective phase conjugation circuit with {RX} array factor combination ability,'' in \emph{2011 IEEE MTT-S International Microwave Symposium}.\hskip 1em plus 0.5em minus 0.4em\relax IEEE, 2011, pp. 1--4.

\bibitem{1456254}
W.~Lindsey and C.~M. Chie, ``A survey of digital phase-locked loops,'' \emph{Proceedings of the IEEE}, vol.~69, no.~4, pp. 410--431, 1981.

\bibitem{9921326}
Y.~Kang, X.~Q. Lin, Y.~Li, and B.~Wang, ``Dual-frequency retrodirective antenna array with wide dynamic range for wireless power transfer,'' \emph{IEEE Antennas and Wireless Propagation Letters}, vol.~22, no.~2, pp. 427--431, 2023.

\bibitem{9955558}
R.~S. Hao, J.~F. Zhang, S.~C. Jin, D.~G. Liu, T.~J. Li, and Y.~J. Cheng, ``{K-/Ka}-band shared-aperture phased array with wide bandwidth and wide beam coverage for {LEO} satellite communication,'' \emph{IEEE Transactions on Antennas and Propagation}, vol.~71, no.~1, pp. 672--680, 2023.

\bibitem{sandhu2016radiating}
A.~I. Sandhu, E.~Arnieri, G.~Amendola, L.~Boccia, E.~Meniconi, and V.~Ziegler, ``Radiating elements for shared aperture tx/rx phased arrays at k/ka band,'' \emph{IEEE Transactions on Antennas and Propagation}, vol.~64, no.~6, pp. 2270--2282, 2016.

\bibitem{xu2019research}
L.~Xu, Y.~Wan, and D.~Yu, ``Research of dual-band dual circularly polarized wide-angle scanning phased array,'' in \emph{2019 IEEE 2nd International Conference on Automation, Electronics and Electrical Engineering (AUTEEE)}.\hskip 1em plus 0.5em minus 0.4em\relax IEEE, 2019, pp. 22--25.

\bibitem{guo2025integrated}
Y.~Guo, S.~Xia, M.~Xiong, Q.~Liu, S.~Zhou, W.~Fang, Q.~Jiang, G.~Yan, and J.~Mu, ``Integrated sensing and communication system based on radio frequency resonance beam,'' \emph{arXiv preprint arXiv:2501.18878}, 2025.

\bibitem{10636970}
Y.~Guo, Q.~Jiang, M.~Xu, W.~Fang, Q.~Liu, G.~Yan, Q.~Yang, and H.~Lu, ``Resonant beam enabled {DoA} estimation in passive positioning system,'' \emph{IEEE Transactions on Wireless Communications}, vol.~23, no.~11, pp. 16\,290--16\,300, 2024.

\bibitem{guo2024resonant}
Y.~Guo, M.~Xiong, W.~Fang, Q.~Jiang, M.~Xu, Q.~Liu, and G.~Yan, ``Resonant beam enabled passive {3D} positioning,'' \emph{IEEE Internet of Things Journal}, pp. 1--1, 2025.

\bibitem{tai1961definition}
C.~Tai, ``On the definition of the effective aperture of antennas,'' \emph{IRE Transactions on Antennas and Propagation}, vol.~9, no.~2, pp. 224--225, 1961.

\bibitem{yu2010aperture}
A.~Yu, F.~Yang, A.~Z. Elsherbeni, J.~Huang, and Y.~Rahmat-Samii, ``Aperture efficiency analysis of reflectarray antennas,'' \emph{Microwave and Optical Technology Letters}, vol.~52, no.~2, pp. 364--372, 2010.

\bibitem{ma2002electromagnetic}
M.~Ma, E.~Larsen, and M.~Crawford, ``Electromagnetic fields with arbitrary wave impedances generated inside a {TEM} cell,'' \emph{IEEE transactions on electromagnetic compatibility}, vol.~33, no.~4, pp. 358--362, 2002.

\end{thebibliography}

\end{document}